\documentstyle[prl,floats,aps,epsf,epsfig]{revtex}

\begin{document}

\draft

\twocolumn[\hsize\textwidth\columnwidth\hsize\csname
@twocolumnfalse\endcsname

\title{General Relativistic Initial Data for the Binary Black Hole /
       Neutron Star System in Quasicircular Orbit }

\author{
Mark~Miller \medskip
}

\address{
McDonnell Center for the Space Sciences \\
Department of Physics,
Washington University, St. Louis, Missouri 63130}

\date{\today}

\maketitle

\begin{abstract}
We present an algorithm for solving the general relativistic initial
value equations for a corotating polytropic star in quasicircular 
orbit with a nonspinning black hole.  The algorithm is used to 
obtain initial data for cases where the black hole mass is 1, 3, and 10 
times larger than the mass of the star.  By analyzing sequences of 
constant baryon mass, constant black hole mass 
initial data sets and carefully monitoring the numerical error,
we find innermost stable circular orbit (ISCO) configuration for these cases. 
While these quasiequilibrium, conformally flat sequences
of initial data sets are not true solutions 
of the Einstein equations (each set,
however, solves the full initial value problem), and thus, we do not
expect the ISCO configurations found here to be completely consistent with 
the Einstein equations, they will be used as convenient starting points
for future numerical evolutions of the full 3+1 Einstein equations.

\end{abstract}


\vskip2pc]

\narrowtext


\section{Introduction}
\label{sec:Introduction}

The inspiraling black hole (BH) / neutron star (NS) binary system is thought
to be both a promising candidate for the central engines of gamma-ray
bursts (GRBs) and a likely source of gravitational waves detectable by
ground-based laser interferometric gravitational wave detectors
(LIGO/VIRGO/TAMA/GEO) or space-based interferometers (LISA).
Theoretical estimates of the event rate of BH/NS mergers~\cite{Narayan91}
give $\sim$ $10^{-6}$ per year per galaxy.  For the advanced detectors of 
LIGO (sensitive out to 1000 Mpc), this would amount
to approximately $1$ event per day.  It is quite likely that 
LIGO-II will be able to extract information about the equation
of state of the nuclear matter inside a neutron star from
the gravitational waves of BH/NS mergers\cite{Vallisneri00}.
In addition, hard GRBs with relatively short duration~\cite{Mao94} could
be produced from BH/NS mergers~\cite{Meszaros00,Paczynski91,Narayan92}.

Apart from the fact that the BH/NS binary system is an interesting
problem in and of itself, a better understanding of the details of BH/NS
mergers is important for observational general relativistic astrophysics.
While some aspects of the BH/NS merger have been investigated with 
Newtonian simulations~\cite{Kluzniak98a,Janka99a}, to date, the complexity
of both the Einstein equations and the relativistic hydrodynamical
equations have prevented studies involving full general relativistic
simulations of BH/NS mergers.  
Here, we present the first steps in the
direction of a fully general relativistic treatment of the coalescence of
an inspiraling binary BH/NS, namely, the calculation of general relativistic 
initial data corresponding to a BH and a quasiequilibrium NS in
a quasicircular orbit, near the innermost stable circular orbit (ISCO).
In contrast to Newtonian theory, initial data in general relativity is
not arbitrary in that one does not have complete freedom in the 
initial choices of the dynamical fields.  The gravitational field
(specified by the 3-metric) and the matter fields must obey 
the Hamiltonian and momentum constraints that take the form
of four coupled, elliptic,
partial differential equations.  We present an algorithm for
numerically solving these constraint equations, together with conditions
that specify initial data corresponding to a quasiequilibrium 
polytropic neutron star
in a corotational orbit with a black hole.  While such a treatment has
been carried out for NS/NS~\cite{Baumgarte98b,Mathews00,Gourgoulhon01} 
and BH/BH systems~\cite{Cook94,Baumgarte00a},
this is the first such treatment for a general relativistic BH/NS system.
In addition, we find configurations that correspond to
an approximate innermost stable circular orbit for the BH/NS system
by defining an effective binding energy, and locating the minimum of 
this binding energy for constant rest mass, constant BH mass
initial data sequences.
Although these are only approximations to the true ISCO configurations
(since a time series of these constant rest mass, constant BH mass 
sequences are not actually 
solutions to the Einstein equations), they will provide a place to begin
studies in a full 3D numerical general relativistic treatment.

The remainder of this paper is outlined as follows.  
In Section~\ref{sec:Equations}, the equations and all assumptions are
presented.  In Section~\ref{sec:Algorithm}, the numerical algorithm
for solving the equations for the BH/NS system is presented.
In Section~\ref{sec:Validation} we present a method of testing the
code to guarantee that we are solving all equations correctly.
In Section~\ref{sec:Sequences} we define an effective 
binding energy and calculate this binding energy
for sequences of constant rest mass, constant BH mass initial data sets.
The data set that 
corresponds to a minimum (when it exists) of the binding energy is taken to be 
an approximate ISCO configuration.
We briefly summarize our results in Section~\ref{sec:Conclusion}.


\section{Equations}
\label{sec:Equations}

The equations one must solve to specify initial data for general relativistic 
systems are the full 4D Einstein equations
\begin{equation}
G_{ab} = 8 \pi T_{ab}
\end{equation}
(here, we are using units where the gravitational constant G and the
speed of light c are set equal to 1),  projected into a direction
normal to a spatial
Cauchy surface (we assume, as always in numerical 
relativity, the entire spacetime to be globally hyperbolic).  
These are the Hamiltonian and momentum constraints, given
respectively by
\begin{equation}
{}^{(3)}R + K^2 - K^{ab} K_{ab} = 16 \pi T_{ab} n^{a} n^{b}
\label{eq:ham_cov}
\end{equation}
and
\begin{equation}
{\cal D}_a K^{ab} - {\gamma}^{ab} {\cal D}_a K = 8 \pi T_{ac} n^{a}
   {\gamma}^{bc},
\label{eq:mom_cov}
\end{equation}
which are elliptic conditions on the initial 3-metric ${\gamma}_{ab}$
and extrinsic curvature $K_{ab}$.
Here, ${}^{(3)}R$ is the scalar curvature of the 3-metric, K is the 
trace of the extrinsic curvature $K_{ab}$, $n^{a}$ is the future
directed unit vector normal
to the Cauchy surface, 
and ${\cal D}$ is the covariant derivative operator compatible with
the 3-metric ${\gamma}_{ab}$.  While it is not necessary to 
specify the lapse and shift functions as part of the initial data,
we find it convenient to do so here.  We take the conditions
on the lapse $\alpha$ and shift vector ${\beta}^a$ to be, 
respectively,
\begin{equation}
{\cal L}_t K = 0
\label{eq:maxslice_cov}
\end{equation}
and
\begin{equation}
{\cal D}^a {\Sigma}_{ab} = 0
\label{eq:mindist_cov}
\end{equation}
where ${\Sigma}_{ab}$ is the distortion tensor, defined by
\begin{equation}
{\Sigma}_{ab} \equiv {\cal L}_t {\gamma}_{ab} -
   \frac {1}{3} {\gamma}_{ab} {\gamma}^{cd} {\cal L}_t {\gamma}_{cd}.
\end{equation}
Eq.~\ref{eq:maxslice_cov} is the 
maximal slicing condition for the lapse, while Eq.~\ref{eq:mindist_cov}
is the 
minimal distortion
equation~\cite{York79} for the shift.

We now make several simplifying assumptions on the form of the metric
${\gamma}_{ab}$, extrinsic curvature $K_{ab}$, and matter variables that
make up the matter stress energy tensor $T_{ab}$.  First, introduce
Cartesian spatial coordinates $\{x^i\}$, where the Latin indices
vary from 1 to 3.  We assume the spatial 3-metric to be conformally
flat
\begin{equation}
\gamma_{ij} = {\phi}^4 {\delta}_{ij}
\end{equation}
and the trace of the extrinsic curvature to vanish
\begin{equation}
K \equiv K_{ij} {\gamma}^{ij} = 0.
\end{equation}
For convenience, we introduce the conformal tracefree extrinsic curvature
${\tilde{A}}^{ij}$ to be 
\begin{equation}
{\tilde{A}}^{ij} = {\phi}^{10} K^{ij}.
\end{equation}
Note that we raise and lower conformal quantities with the conformal
(flat) metric ${\delta}_{ij}$, so that, e.g., 
${\tilde{A}}_{ij} = {\phi}^2 K_{ij}$.
Furthermore, we assume the stress energy tensor to be that of a 
perfect fluid, namely,
\begin{equation}
T^{ab} = {\rho}_0 h u^a u^b + P g^{ab}
\end{equation}
where ${\rho}_0$ is the rest mass density of the fluid, 
$u^a$ is the 4-velocity of 
the fluid, $P$ is the pressure of the fluid, 
$g^{ab}$ is the 4-metric, and $h$ is the 
relativistic specific enthalpy given by
\begin{equation}
h = 1 + \epsilon + P/{\rho}_0
\end{equation}
where $\epsilon$ is the specific internal energy density of the fluid.
For the rest of this paper, we assume the perfect fluid equation
of state, along with the polytropic equation of state:
\begin{equation}
P = (\Gamma - 1) {\rho}_0 \epsilon = k {{\rho}_0}^{\Gamma}.
\label{eq:poly_eos}
\end{equation}
where $\Gamma$ is the adiabatic index of the fluid (all numerical results
for this paper use $\Gamma = 2$) and $k$ is the polytropic
constant of the fluid.  For convenience, we completely fix the units by
setting the polytropic constant $k$, along with $G$ and $c$, equal to 1.
We define the 3-velocity $v^i$ with respect
to velocity of the coordinates $\{x^i\}$, and note that these components 
are related to the spatial components of the 4-velocity, $u^i$, by the
following:
\begin{equation}
u^i = W (v^i - {\beta}^i / \alpha)
\end{equation}
where W is the Lorentz factor, 
$W = \alpha u^t = {(1 - {\gamma}_{ij} v^i v^j)}^{-1/2}$.

\subsection{General Relativistic Equations}

There are several ways to accommodate the presence of a black hole
in the computational domain for initial data 
preparation, including conformal imaging with horizon
excision~\cite{Cook93}.  We adopt a method
given by Brandt and Br{\"u}gmann~\cite{Brandt97b}, where
the singular nature of the black hole is taken explicitly into
account by rewriting the conformal factor $\phi$ in terms of the
function $\chi$ as
\begin{equation}
\phi = \frac {M_{BH}}{ 2 \; |\vec{x} - {\vec{x}}_{BH}| } + \chi ,
\label{eq:conf_fact}
\end{equation}
where $M_{BH}$ is the bare mass of the black hole 
puncture (see~\cite{Brandt97b})
and ${\vec{x}}_{BH}$
is the coordinate location of the black hole puncture.
The Hamiltonian constraint, Eq.\ref{eq:ham_cov}, is now written as
\begin{equation}
{\partial}_i {\partial}^i \chi + \frac {1}{8} {\phi}^{-7} 
{\tilde{A}}_{ij} {\tilde{A}}^{ij} + 2 \pi {\phi}^{5} ({\rho}_0 h W^2 - P) = 0.
\label{eq:ham_nohat}
\end{equation}
As with the conformal factor, we rewrite the conformal tracefree
extrinsic curvature ${\tilde{A}}^{ij}$ in terms of the functions
${\tilde{Z}}^{ij}$ as
\begin{eqnarray}
{\tilde{A}}^{ij} & = & {\tilde{Z}}^{ij} + \frac {3}{2 \; {|\vec{x} - 
{\vec{x}}_{BH}|}^2 } * \nonumber \\
 &  & \left ( P^i_{BH} n^j_{BH} + P^j_{BH} n^i_{BH}  \right.  \nonumber \\
 &  &  \left.   - ({\delta}^{ij} - n^i_{BH} n^j_{BH}) 
P^k_{BH} {n_k}_{BH} \right ) 
\label{eq:a_define}
\end{eqnarray}
where $P^i_{BH}$ is the linear momentum of the black hole as measured 
from infinity, and $n^i_{BH} = (x^i - x^i_{BH})/|\vec{x} - {\vec{x}}_{BH}|$.
The momentum constraint, Eq.~\ref{eq:mom_cov}, is now written as 
\begin{equation}
{\partial}_j {\tilde{Z}}^{ij} - 8 \pi {\phi}^{10} {\rho}_0 h W^2 v^i = 0.
\label{eq:mom_nohat}
\end{equation}
Making the ansatz
\begin{equation}
{\tilde{Z}}^{ij} = \frac { {\chi}^7 } {2 \eta}
   \left ( {\partial}^i W^j + {\partial}^j W^i -
   \frac{2}{3} {\delta}^{ij} {\partial}_k W^k \right )
\label{eq:mom_ansatz}
\end{equation}
giving the six functions ${\tilde{Z}}^{ij}$ in terms of the three 
functions $W^i$, the three momentum constraints can be written as 
equations for the three functions $W^i$:
\begin{eqnarray}
{\partial}_j {\partial}^j W^i + 
\frac{1}{3} {\partial}^i {\partial}_j W^j - 
   {\tilde{Z}}^{ij} {\partial}_j \frac{2 \eta}{ {\chi}^7} & & \nonumber \\
 -  16 \pi \eta {\phi}^{10} {\chi}^{-7} {\rho}_0 h W^2 v^i & = & 0.
\label{eq:mom_ansatz_nohat}
\end{eqnarray}

Using the Einstein equations, the
maximal slicing condition, Eq.~\ref{eq:maxslice_cov}, can be written
in terms of the function $\eta \equiv \alpha \phi$ as
\begin{eqnarray}
{\partial}_i {\partial}^i \eta
    & + & \eta
    \left (  - \frac {7}{8} {\phi}^{-8} {\tilde{A}}_{ij} {\tilde{A}}^{ij} 
      + 2 \pi {\phi}^{4} (2 {\rho}_0 h - 3 {\rho}_0 h W^2 - 5 P) 
   \right )  \nonumber \\
 & = & 0.
\label{eq:maxslice_nohat}
\end{eqnarray}
The minimal distortion shift equation, Eq.~\ref{eq:mindist_cov} can
be written
\begin{eqnarray}
  {\partial}_j {\partial}^j {\beta}^i & + &
   \frac {1}{3} {\partial}_j {\partial}^i {\beta}^j =  \nonumber \\
 & & \mbox{} - \frac {6}{\phi}
   \left ( ({\partial}^j {\beta}^i + {\partial}^i {\beta}^j){\partial}_j \phi -
      \frac {2}{3} ({\partial}^i \phi) 
                 {\partial}_j {\beta}^j \right ) \nonumber \\
 & &  \mbox{} + 2 {\phi}^{-6} {\tilde{A}}^{ij} {\partial}_j \alpha 
   + 16 \pi \alpha {\phi}^4 {\rho}_0 h W^2 v^i .
\end{eqnarray}
Notice that the ansatz for the form of the extrinsic curvature,
Eqs.~\ref{eq:a_define}~and~\ref{eq:mom_ansatz}, is not equivalent to
that used in NS/NS studies~\cite{Baumgarte98b,Mathews00,Gourgoulhon01}.
Due to the presence of a BH in the form of Eq.~\ref{eq:conf_fact} with
non-zero momentum~\cite{Brandt97b}, we cannot put the extrinsic 
curvature in the form (analogous to Eq.~\ref{eq:mom_ansatz})
\begin{equation}
{\tilde{A}}^{ij} = \frac { {\phi}^6 } {2 \alpha}
   \left ( {\partial}^i W^j + {\partial}^j W^i -
   \frac{2}{3} {\delta}^{ij} {\partial}_k W^k \right )
\label{eq:dt0_mom_ansatz}
\end{equation}
which is the condition that the time derivative of the conformal
metric vanish, namely, ${\cal{L}}_t ({\phi}^{-4} {\gamma}_{ij}) = 0$
(in this case, the vector potential $W^i$ assumes the role of the
shift). 
However, Eqs.~\ref{eq:a_define}~and~\ref{eq:mom_ansatz} reduce to 
Eq.~\ref{eq:dt0_mom_ansatz} as $M_{BH} \rightarrow 0$.

\subsection{Matter Equations}

In writing down the equations that govern the matter, we follow closely
the formalism of~\cite{Baumgarte98b}, where corotating 
binary neutron stars were 
solved in the quasiequilibrium approximation.  First, we assume that
the 4-velocity vector field of the fluid $u^{a}$ is proportional to
an approximate Killing vector field, and write the 4-velocity as
\begin{equation}
u^a = u^t \left ( \alpha n^a + {\beta}^a + \Omega 
   {\left ( \frac {\partial}{\partial \phi} \right ) }^a \right )
\end{equation}
(recall that the vector field representing the flow of time is
$t^a = \alpha n^a + {\beta}^a$), where $\Omega$ is the (constant) angular
velocity.  The normalization condition on the
4-velocity, $u^a u_a = -1$, now becomes
\begin{equation}
v^2 \equiv {\gamma}_{ij} v^i v^j = \frac { { \left ( \vec{\beta} + \Omega 
   \vec{\left ( \frac {\partial}{\partial \phi} \right ) } \right ) }^2}
   {{\alpha}^2}
\end{equation}
The relativistic Bernoulli equation, under the assumption that the
EOS is of the form of Eq.~\ref{eq:poly_eos}, can be directly 
integrated to yield
\begin{equation}
\frac{u^t}{h} = \verb+constant+
\end{equation}
which can now be written as
\begin{equation}
h \sqrt{ {\alpha}^2 - {\phi}^4 
   \left ( {( {\beta}^x - y \Omega )}^2 + {( {\beta}^y + x \Omega )}^2 +
      {( {\beta}^z )}^2 \right ) } = C_B,
\label{eq:bernoulli}
\end{equation}
where $C_B$ is a constant.  


\section{Numerical Algorithm}
\label{sec:Algorithm}

We will now describe the numerical algorithm used to solve the equations
in the previous section.  What we require are simultaneous solutions
of the 9 equations (the Hamiltonian constraint (1),  the maximal slicing 
condition (1), the momentum constraints (3), the minimal distortion shift 
equations (3), and the relativistic Bernoulli equation (1)) for the 
9 fields $\chi$, $W^i$, $\eta$, ${\beta}^i$, and ${\rho}_0$, which 
completely specify the configuration.  There are
3 free parameters that completely specify the 
configuration of a neutron star and black hole in quasicircular
orbit about each other.  These will be the the ratio of the 
mass of the black hole to the mass of the neutron star (${\mu}_{bn}$), 
the separation between the black hole and the neutron star (parametrized
by the parameter ${\hat{x}}_A$, which is defined below), 
and the maximum rest mass density ${({\rho}_0)}_{max}$ of the neutron
star.  The idea is to iteratively solve the required equations with
a stable iterative procedure in which the residual of 
each equation decreases to some desired tolerance.   To this end, we
start by following~\cite{Baumgarte98b} in defining new spatial
coordinates $\{{\hat{x}}^i\}$ which are just a rescaling of the 
original coordinates $\{x^i\}$:
\begin{equation}
{\hat{x}}^i = \frac { x^i } {\sigma}.
\end{equation}
We transform all quantities (except for the matter fields) to these new
coordinates $\{{\hat{x}}^i\}$.  All numerical computations are done
in these hatted coordinates.  We can easily transform any calculated 
quantity back to the unhatted coordinates, e.g., 
${\partial}_i =  {\hat{\partial}}_i / \sigma$, 
$ {\tilde{A}}^{ij} = {\hat{\tilde{A}^{ij}}} / \sigma$, 
$ \Omega = {\hat{\Omega}} / \sigma$, or 
$ M_{BH} = \sigma {\hat{M}}_{BH}$.

We choose for the black hole and the neutron star to be orbiting in the 
$\hat{z} = 0$ plane, with their respective centers lying on the 
$\hat{x}$-axis.  The neutron star surface will intersect the $\hat{x}$-axis 
at two different points.  We define the intersection furthest from the 
origin of the coordinate system to be exactly $1$, and denote the 
intersection closest to the origin as ${\hat{x}}_A$.  
We require the 
bare mass of the black hole $\hat{M}_{BH}$ to be given in terms of the
ADM mass of the neutron star
\begin{equation}
\hat{M}_{BH} = {\mu}_{bn} \; {(\hat{M}_0)}_{ADM}
\label{eq:bh_mass}
\end{equation}
where ${(\hat{M}_0)}_{ADM}$ denotes the ADM mass of a neutron star 
in isolation with
total rest mass ${\hat{M}_0}$.
We further require the center of mass of the
system to be at the origin of the coordinate system.  We therefore set the
coordinate of the black hole puncture $\hat{x}^i_{BH} = (\hat{x}_{BH},0,0)$
to be 
\begin{equation}
\hat{x}_{BH} = - \frac {\hat{x}_C}{{\mu}_{bn}}
\label{eq:cm_prescript}
\end{equation}
where $\hat{x}_C$ is the coordinate value along the $\hat{x}$-axis 
where the rest mass density ${\rho}_0$ of the neutron star acquires
its maximum value.  As $\hat{x}_C$ will generally not be located on a 
discrete grid point, a quadratic polynomial is fit to the maximum 
discrete value and its nearest discrete neighbors in the $\hat{x}$-direction.
It is the coordinate 
location of the maximum of this polynomial that is denoted as $\hat{x}_C$.
One may worry that the prescription for the location of the black hole
along the $\hat{x}$-axis, Eq.~\ref{eq:cm_prescript}, may not actually place
the center of mass of the system precisely at the origin of the
coordinate system.  Analytically, the center of mass of the 
system can only be determined by observers at infinity.  In our 
case, one only needs to look at the
$1/\hat{r}$ falloff of the conformal factor $\phi$ along the 
$\hat{x}$-axis.  This is particularly easy in our case, as the conformal
factor is written explicitly in terms of a $1/{\hat{r}}_{BH}$ piece 
(${\hat{r}}_{BH} = |{\hat{x}}^i - {{\hat{x}}^i}_{BH}|$ is the
coordinate distance from the black hole
puncture) and $\chi$, which will be numerically solved with a robin boundary
condition of $(\chi - 1) \; \sim \; 1/{\hat{r}}_{NS}$, where
${\hat{r}}_{NS} = |{\hat{x}}^i - {(\hat{x}_C,0,0)}^i|$.
One could alternatively define the ADM mass of
the neutron star ${(\hat{M}_0)}_{ADM}$ in Eq.~\ref{eq:bh_mass} in terms of
the $1/{\hat{r}}_{NS}$ falloff of the function $\chi$, and one could argue that
this would be a more accurate way of restricting the center of mass of
the system to be at the origin of the coordinate system.  We have checked
that both methods of defining the ADM mass of the neutron star are the
same to within 6\% at
worst, and much better than this in most cases.  

We determine the momentum of the black hole, 
$\hat{P}^i_{BH} = (0,\hat{P}^{\hat{y}}_{BH},0)$, from the condition that the 
total linear momentum in the $\hat{y}$-direction vanish.  The 
linear momentum in the $\hat{y}$-direction is 
given (see, e.g., \cite{Wald84}) by 
\begin{equation}
\hat{P}^{\hat{y}} = \frac {1}{8 \pi} \lim_{ \hat{r} \rightarrow \infty} 
   \int d^2\hat{S}_i \; (\hat{K}^{\hat{y}i} - \hat{K}{\delta}^{\hat{y}i})
\label{eq:bh_mom}
\end{equation}
For the case of conformally flat initial data, where the 
conformal tracefree extrinsic curvature is written as 
in Eq.~\ref{eq:a_define}, the condition that the total linear
momentum in the $\hat{y}$-direction $\hat{P}^{\hat{y}}$ vanish is 
simply 
\begin{equation}
\hat{P}^{\hat{y}}_{BH} = - \int d^3\hat{x} \; {\sigma}^2 {\phi}^{10}
   {\rho}_0 h W^2 v^{y},
\label{eq:phyhat}
\end{equation}
where the momentum constraint Eq.~\ref{eq:mom_nohat} has been used to
simplify the integral.

The form of the 
Hamiltonian constraint (Eqs.~\ref{eq:ham_cov},\ref{eq:ham_nohat}), 
momentum constraints (Eqs.~\ref{eq:mom_cov},\ref{eq:mom_ansatz_nohat}), 
and maximal slicing condition 
(Eqs.~\ref{eq:maxslice_cov},\ref{eq:maxslice_nohat})
that we solve numerically can now
be written, respectively, as
\begin{equation}
{\hat{\partial}}_i {\hat{\partial}}^i \chi + \frac {1}{8} {\phi}^{-7}
{\hat{\tilde{A}}}_{ij} {\hat{\tilde{A}}}^{ij} + 
2 \pi {\sigma}^2 {\phi}^{5} ({\rho}_0 h W^2 - P) = 0,
\label{eq:ham_hat}
\end{equation}
\begin{eqnarray}
{\hat{\partial}}_j {\hat{\partial}}^j {\hat{W}}^i & + &
\frac{1}{3} {\hat{\partial}}^i {\hat{\partial}}_j {\hat{W}}^j \nonumber \\ 
 & & \mbox{} -  {\hat{\tilde{Z}}}^{ij} {\hat{\partial}}_j 
                   \frac{2 \eta}{ {\chi}^7} -
   16 \pi {\sigma}^2 \eta {\phi}^{10} {\chi}^{-7} {\rho}_0 h W^2 v^i = 0,
\label{eq:mom_ansatz_hat}
\end{eqnarray}
and
\begin{eqnarray}
{\hat{\partial}}_i {\hat{\partial}}^i \eta & + & \eta
   \left ( - \frac {7}{8} 
      {\phi}^{-8} {\hat{\tilde{A}}}_{ij} {\hat{\tilde{A}}}^{ij} +
      2 \pi {\sigma}^2 {\phi}^{4} (2 
         {\rho}_0 h - 3 {\rho}_0 h W^2 - 5 P) \right ) \nonumber \\
 & & = 0.
\label{eq:maxslice_hat}
\end{eqnarray}
In hatted coordinates, the conformal factor is written as
\begin{equation}
\phi = \frac {\hat{M}_{BH}}{ 2 \; |\hat{\vec{x}} - {\hat{\vec{x}}}_{BH}| } + 
   \chi ,
\end{equation}
whereas the momentum constraint ansatz, Eq.~\ref{eq:mom_ansatz}, is written
as 
\begin{equation}
{\hat{\tilde{Z}}}^{ij} = \frac { {\chi}^7 } {2 \eta}
   \left ( {\hat{\partial}}^i \hat{W}^j + {\hat{\partial}}^j \hat{W}^i -
   \frac{2}{3} {\delta}^{ij} {\hat{\partial}}_k \hat{W}^k \right ).
\label{eq:ansatz_hat}
\end{equation}

In numerically solving the matter equations (the relativistic Bernoulli
equation, Eq.~\ref{eq:bernoulli}), written in
hatted coordinates here as
\begin{equation}
h \sqrt{ {\alpha}^2 - {\phi}^4
   \left ( {( {\beta}^x - \hat{y} \hat{\Omega} )}^2 + 
      {( {\beta}^y + \hat{x} \hat{\Omega} )}^2 +
      {( {\beta}^z )}^2 \right ) } = C_B,
\label{eq:bernoulli_hat}
\end{equation}
we again follow~\cite{Baumgarte98b}
by choosing three points along the $\hat{x}$-axis, and evaluating
Eq.~\ref{eq:bernoulli_hat} at these three points.  Using Newton's method,
we solve these three equations for the three constants $\sigma$, 
$\hat{\Omega}$, and $C_B$.  
Using these three newly computed constants,
we reset the matter variable ${\rho}_0$ everywhere through
Eq.~\ref{eq:bernoulli_hat}.  This uniquely specifies all of the 
matter variables $h$, $v^i$, and $W$.  
The three points we evaluate Eq.~\ref{eq:bernoulli_hat} are where the
surface of the neutron star intersects the $\hat{x}$-axis 
($\hat{x} = 1, \hat{x}_A$), and where 
${\rho}_0$ attains its maximum value ($\hat{x} = \hat{x}_C$).
Solving Eq.~\ref{eq:bernoulli_hat} at three points for the 
constants $\sigma$, $\hat{\Omega}$, and $C_B$ is not straightforward.  
While $\hat{\Omega}$ and $C_B$ appear explicitly in Eq.~\ref{eq:bernoulli_hat},
$\sigma$ does not.  However, the lapse $\alpha$ 
and the conformal factor $\phi$
depend on $\sigma$, as can be seen 
from Eqs.~\ref{eq:ham_hat}~and~\ref{eq:maxslice_hat}.  In both of 
these equations, increasing (decreasing) $\sigma$ acts like 
an increase (decrease) in the 
matter source, thus resulting in a decrease (increase) in the lapse $\alpha$, 
and an increase (decrease)
in the conformal factor $\phi$ in the vicinity of the NS.  
In~\cite{Baumgarte98b}, the functions
$\alpha = \alpha(\sigma)$ and $\phi = \phi(\sigma)$ were modeled using
Newtonian scaling relations, which we found in our case to not produce
a stable (converging) numerical algorithm for iteratively solving the
differential spacetime equations with the algebraic relativistic
Bernoulli's equation.  Instead, we use static spherical solutions
of the Einstein's equations coupled to a perfect fluid 
(Tolman-Oppenheimer-Volkoff~\cite{Tolman39,Oppenheimer39b}, or TOV, solutions)
to model the functions $\alpha = \alpha(\sigma)$ and $\phi = \phi(\sigma)$.
For example, at $\hat{x}=\hat{x}_C$, we take $\alpha_C(\sigma)$, the function
we will use to model the dependence of the lapse $\alpha$ on the 
scaling factor $\sigma$ at the 
point of maximum rest mass density of the neutron star, to be 
\begin{equation}
\alpha_C(\sigma) = {\alpha}_0 + {\alpha}_{TOV}(\sigma)
\label{eq:alp_model}
\end{equation}
where ${\alpha}_{TOV}(\sigma)$ is the lapse at the center of the TOV star with 
scaling factor $\sigma$, and ${\alpha}_0$ is a constant specified by the 
consistency condition that the 
actual value of the lapse at $\hat{x} = \hat{x}_C$ should be given
by the original value of the scaling factor, ${\sigma}_{old}$,
\begin{equation}
{\alpha}_0 \equiv \alpha_C({\sigma}_{old}) - {\alpha}_{TOV}({\sigma}_{old})
\end{equation}
Similar functions are used to model the $\sigma$ dependence of the
lapse function $\alpha$ at the surface of the stars ($\hat{x} = \hat{x}_A,1$),
as well as for the conformal factor $\phi$.  We can now solve 
Eq.~\ref{eq:bernoulli_hat} at the points $\hat{x} = \hat{x}_A, \hat{x}_C, 1$
using Newton's method, 
where the modeling functions (e.g., Eq.~\ref{eq:alp_model}) are substituted
for $\alpha$ and $\phi$.  

The last equation we need to solve numerically is the minimal 
distortion shift equation, Eq.~\ref{eq:mindist_cov}, written
in hatted coordinates as
\begin{eqnarray}
{\hat{\partial}}_j {\hat{\partial}}^j {\beta}^i & + &
   \frac {1}{3} {\hat{\partial}}_j {\hat{\partial}}^i {\beta}^j = \nonumber \\
 & & \mbox{} - 
   \frac {6}{\phi}
   \left ( ({\hat{\partial}}^j {\beta}^i + 
            {\hat{\partial}}^i {\beta}^j){\partial}_j \phi -
      \frac {2}{3} ({\hat{\partial}}^i \phi) 
                    {\hat{\partial}}_j {\beta}^j \right ) \nonumber \\
 & & \mbox{}  + 2 {\phi}^{-6} {\hat{\tilde{A}}}^{ij} {\hat{\partial}}_j \alpha
   + 16 \pi {\sigma}^2 \alpha {\phi}^4 {\rho}_0 h W^2 v^i.
\label{eq:mindist_hat}
\end{eqnarray}
Numerically, we find the first term on the
right hand side of Eq.~\ref{eq:mindist_hat}
(terms involving $(\hat{\partial}\beta)(\hat{\partial}\phi)/\phi$)
quite difficult to handle,
due to the built in singular point of the conformal factor $\phi$ at
the black hole puncture, ${\hat{r}}_{BH} = 0$.  
However, if one writes down the minimal
distortion shift equations in spherical coordinates about the point
$\hat{x}^i_{BH}$, one finds that the only nontrivial (non-constant)
radial solutions to
the equations are of the form $\beta^i \sim \hat{r}_{BH}$ or 
$\beta^i \sim \hat{r}_{BH}^4$ as
the radial coordinate about $\hat{x}^i_{BH}$, 
$\hat{r}_{BH} \equiv |\hat{\vec{x}} - \hat{\vec{x}}_{BH}|$, goes to zero.
We therefore excise the region about the black hole for the minimal distortion
shift equation.  Note that, as the only place the shift appears 
(other than the minimal distortion shift equation) is in
Bernoulli's equation, Eq.~\ref{eq:bernoulli_hat}, which only applies 
where the matter content of the spacetime is nonzero, the minimal
distortion shift equation is the only equation where black hole
excision is needed.  We excise a cubical region of the computational
domain for the minimal distortion shift equation, where the excision cube
is centered about $\hat{x}^i_{BH}$.  The sides of the cube are taken to
have a length of $\hat{M}_{BH}/2$ or $3 \; \Delta \hat{x}$, whichever is 
larger.  We find that we need the cubical excised region to be at least
three grid points wide in order to put a reasonable boundary condition there.
The boundary condition that we use for the shift at the boundary
of the excised region is
\begin{equation}
{\beta}^i = {\beta}_0^i + {\cal{O}}(\hat{r}_{BH})
\end{equation}
where ${\beta}_0^i$ is a constant vector obtained by performing a
Lorentz boost at infinity, with a boost factor corresponding to a velocity
of $\hat{P}_{BH} / \hat{M}_{BH}$, on the Schwarzschild solution written
in isotropic coordinates, as $\hat{r}_{BH} \rightarrow 0$.  Specifically,
we have
\begin{equation}
{\beta}_0^i = \left ( 0,- \frac{ \hat{P}_{BH} / \hat{M}_{BH} }
   {\sqrt{1 - {(\hat{P}_{BH} / \hat{M}_{BH})}^2}},0 \right ).
\end{equation}

The algorithm is initialized by choosing values for the three
parameters ${({\rho}_0)}_{max}$, ${\mu}_{bn}$, and $\hat{x}_A$,
which will remain fixed during the entire algorithm.  We 
have coded a Full Approximation Scheme (FAS) 
multigrid solver~\cite{BrandtA94} to solve 
Eqs.~\ref{eq:ham_hat},~\ref{eq:mom_ansatz_hat},~\ref{eq:maxslice_hat},
and~\ref{eq:mindist_hat} simultaneously, thus obtaining solutions
to the Hamiltonian constraint, momentum constraints, maximal slicing
condition, and minimal distortion shift equations, respectively.
We then solve Bernoulli's equation, Eq.~\ref{eq:bernoulli_hat}, 
at the three points $\hat{x} = 1,{\hat{x}}_A, {\hat{x}}_C$ for 
the three constants $\sigma$,
${\hat{\Omega}}$, and $C_B$.  These 3 constants are then used
to set the matter field ${\rho}_0$ using 
Bernoulli's equation Eq.~\ref{eq:bernoulli_hat}.  We then 
set the constants $\hat{M}_{BH}$, $\hat{P}_{BH}$, and $\hat{x}_{BH}$ via.
the prescription 
Eqs.~\ref{eq:bh_mass},~\ref{eq:phyhat},~and~\ref{eq:cm_prescript}, 
respectively.  We repeat the above process until the entire 
configuration has converged to a simultaneous solution.  
Typically, we stop the iteration
process when the constants $\sigma$, ${\hat{\Omega}}$, 
and $C_B$ change by less than one part in 10,000 from one
iteration to the next.


\section{Code Validation}
\label{sec:Validation}

An important and often overlooked ingredient in numerical relativity
is code validation (e.g., see~\cite{Flanagan99} and references therein).  
Due to the complexity of the Einstein equations, 
it is extremely important to verify that the coded finite difference equations
are consistent with the differential equations one wants to 
solve.  That
is, as the discretization parameter $\Delta$ goes to zero, the finite 
difference equations that are solved in the code (denoted by the
operator equation ${\cal{L}}_\Delta = 0$) should be verified to
approach the differential equations (denoted by the operator 
equation ${\cal{L}} = 0$) as
some integral power of $\Delta$:
\begin{equation}
{\cal{L}}_{\Delta} - {\cal{L}} = {\cal{O}} ({\Delta}^n),
\label{eq:consistent}
\end{equation}
where the integer $n$ is determined by the type of finite differencing used.
Typically, one chooses second-order finite differencing (as
we have done in this paper), in which case $n=2$.  Only when a numerical
code used to solve ${\cal{L}}_\Delta$ is shown to satisfy 
Eq.~\ref{eq:consistent} can that code be considered validated. 

It turns out to
be quite straightforward to verify any numerical solver ${\cal{L}}_\Delta$
for Eq.~\ref{eq:consistent}, using the notion of an Independent 
Residual Evaluator (IRE)~\cite{ChoptuikPrivateComm}.
The idea is to construct, through an independent means (by either writing 
the differential equation in a different form or using different
finite differencing schemes, or preferably both), a residual evaluator
for the differential equations represented by ${\cal{L}}$.  This is usually
much easier to do than to code up a numerical solver for ${\cal{L}}$.  
Denote this IRE of ${\cal{L}}$ to be ${\cal{L}}'_\Delta$, and assume
that the truncation error is of the same order $n$ as that
of the numerical solver of ${\cal{L}}_\Delta$ implemented by the code.  
We now write Eq.~\ref{eq:consistent} as 
\begin{equation}
{\cal{L}}_{\Delta} - {\cal{L}}'_\Delta = {\cal{O}} ({\Delta}^n),
\label{eq:newconsistent}
\end{equation}
We can now compute numerically the 
left hand side of Eq.~\ref{eq:newconsistent}
directly.  Simply use the code to solve the finite difference
equations ${\cal{L}}_\Delta = 0$ and compute the IRE ${\cal{L}}'_\Delta$
on the resulting numerical solution.  By Eq.~\ref{eq:newconsistent}, the
value of this IRE should approach zero as ${\cal{O}} ({\Delta}^n)$.  By
performing this numerical calculation at different resolutions
(i.e., different values of the discretization parameter $\Delta$),
one checks whether or not this is the case.  Only once the IRE is shown to
approach zero as ${\cal{O}} ({\Delta}^n)$ can the code be considered 
to be validated.  

Of course, due to the specific nature of any 
particular numerical solution, errors in a numerical code which solves the
finite difference equations ${\cal{L}}_{\Delta}$ may not become
apparent with simply one particular numerical solution.  This is why
we advocate the verification of Eq.~\ref{eq:newconsistent} not only
with some fiducial numerical solution, but {\it for {\bf all} numerical 
results produced by a code}!

In our case, we use a second order accurate Full Approximation Scheme (FAS)
multigrid method to solve the eight coupled elliptic 
partial differential equations
which are the Hamiltonian constraint (Eq.~\ref{eq:ham_hat}),
the momentum constraints (Eqs.~\ref{eq:mom_ansatz_hat}), the maximal
slicing condition (Eq.~\ref{eq:maxslice_hat}), and the minimal
distortion shift equations (Eqs.~\ref{eq:mindist_hat}).  In addition,
we code an IRE for each equation that we solve numerically.  To guarantee
that the IRE is truly independent of the numerical solver, we use
the covariant forms of the Hamiltonian constraint (Eq.~\ref{eq:ham_cov}),
momentum constraints (Eqs.~\ref{eq:mom_cov}), maximal slicing condition
(Eq.~\ref{eq:maxslice_cov}), and minimal distortion shift equations
(Eqs.~\ref{eq:mindist_cov}), computed in the hatted 
(computational) coordinates $\hat{x}^i$
with absolutely no assumptions on the form of the metric or extrinsic
curvature.  For each numerical solution computed using the process
described in Section~\ref{sec:Algorithm}, we calculate 
the value of each IRE, and verify that it goes to zero as
${\cal{O}}({\Delta}^2)$ (since we have used second-order accurate methods
in both our solvers and our IREs).  

\begin{figure}
\epsfxsize=3in
\begin{center}
\leavevmode
\epsffile{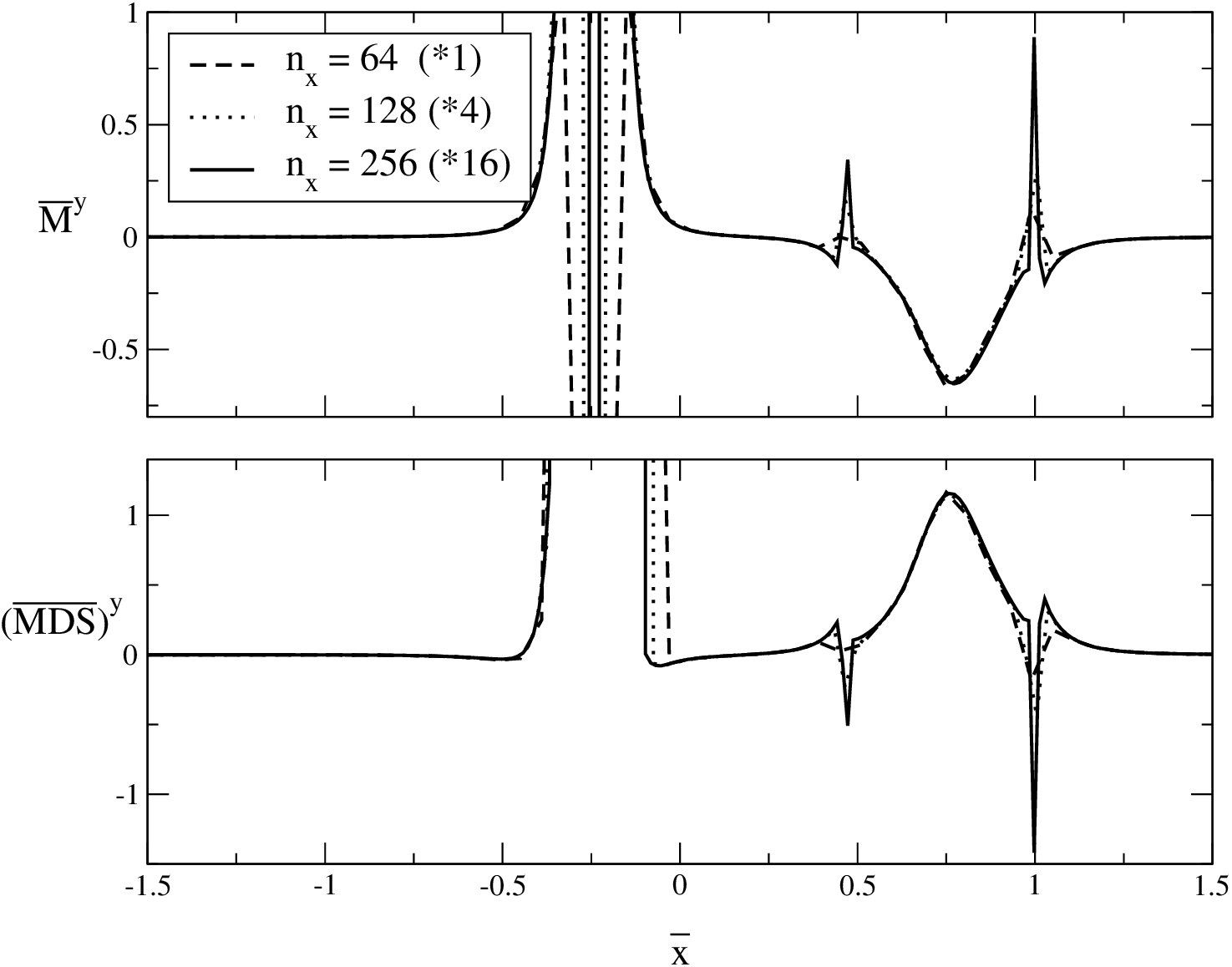}
\end{center}
\caption{ An independent residual evaluator (IRE) evaluated on the numerical
solution obtained for the configuration specified by ${\mu}_{BH} = 3$, 
${\hat{x}}_A = 0.4668$, and ${({\rho}_0)}_{max} = 0.09265$.  Shown is
the value of the independent residual for the $\hat{y}$-component of both the 
momentum constraint ${\overline{M}}^y$ and minimal distortion
shift equation ${\overline{MDS}}^y$ along the $\hat{x}$-axis for 
3 separate resolutions.
}
\label{fig:ire1}
\end{figure}

\begin{figure}
\epsfxsize=3in
\begin{center}
\leavevmode
\epsffile{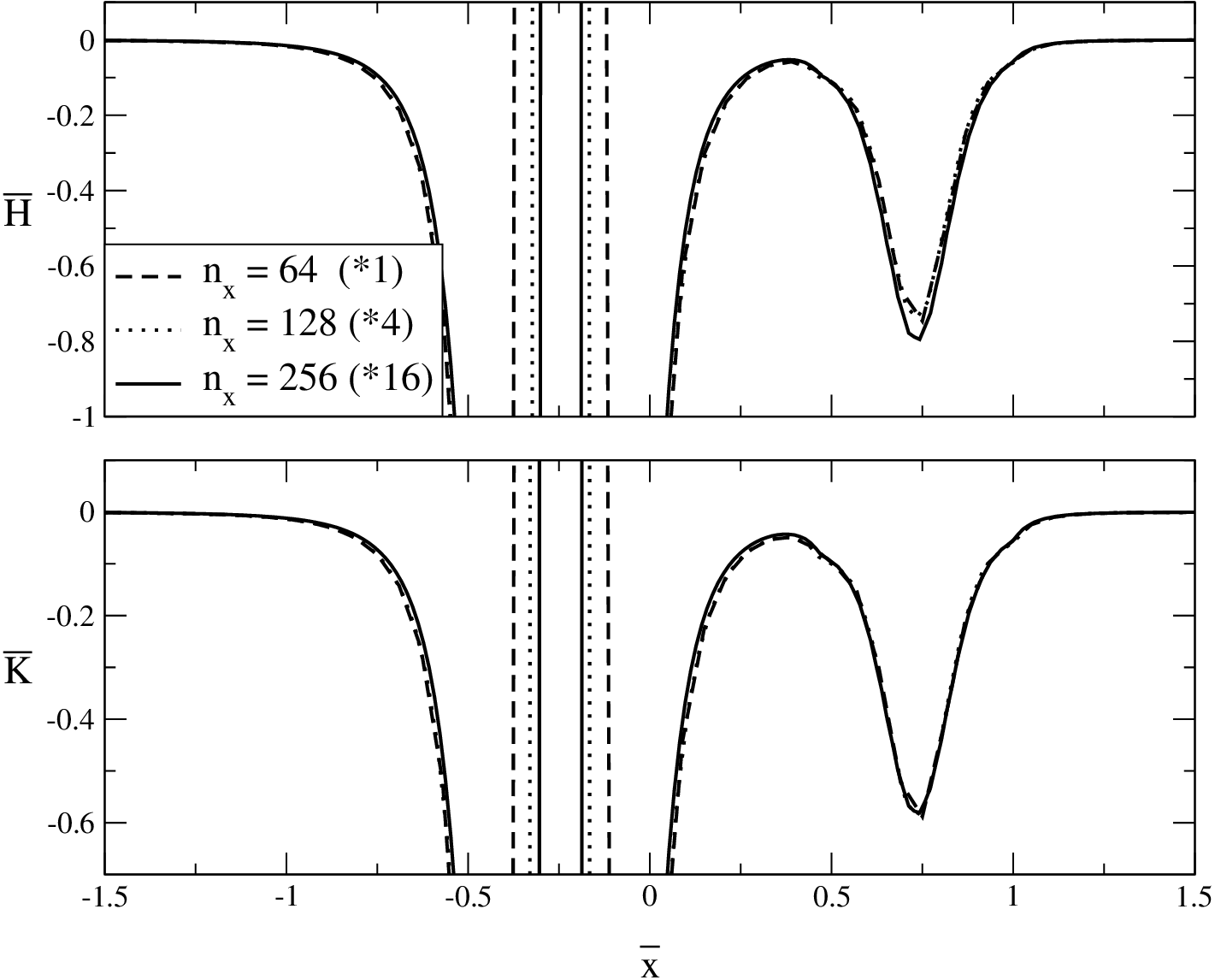}
\end{center}
\caption{ An independent residual evaluator (IRE) evaluated on the numerical
solution obtained for the configuration specified by ${\mu}_{BH} = 3$,
${\hat{x}}_A = 0.4668$, and ${({\rho}_0)}_{max} = 0.09265$.  Shown is
the value of the independent residual for the Hamiltonian constraint
${\overline{H}}$ and maximal slicing 
equation ${\overline{K}}$ along the $\hat{x}$-axis for 
3 separate resolutions.
}
\label{fig:ire2}
\end{figure}

It would not be feasible to show the results of all of the convergence
tests of the IREs which were, in fact, performed on each numerical solution
obtained in compiling the results of this paper.  
Instead I will present an example of just one of the convergence tests of the 
IREs, and state that this same test was performed on all numerical
solutions obtained for this paper.  For this example, we set
the three free parameters as follows:
the ratio of the
mass of the black hole to the mass of the neutron star ${\mu}_{bn} = 3$,
the separation parameter ${\hat{x}}_A = 0.4668$, and the maximum rest
mass density of the neutron star ${({\rho}_0)}_{max} = 0.09265$.  Using the
algorithm described in Section~\ref{sec:Algorithm}, we numerically
solve for the 9 fields $\chi$, $\hat{W}^i$, $\eta$, ${\beta}^i$, and
${\rho}_0$.  This corresponds to a configuration in which the rest mass of the
neutron star $\hat{M}_0$ is approximately 73\% that of the maximum
rest mass of a stable TOV configuration (we have set the adiabatic
index $\Gamma = 2$ for this example).  Using this numerically 
generated solution, we construct the 3-metric, extrinsic curvature,
lapse, and shift.  We then compute the IRE of the 8 partial differential 
equations at each point on the computational grid.  This entire process
is repeated for resolutions of $\Delta \hat{x} = 0.06$, $0.03$, and $0.015$,
using the number of grid points $n_x$ along the $\hat{x}$-axis to be
$n_x = 64$, $128$, and $256$, respectively.
Shown in Figures~\ref{fig:ire1}~and~\ref{fig:ire2}
are the values of the IRE along the $\hat{x}$-axis at the
three resolutions of the $\hat{y}$-component of the momentum constraint,
the $\hat{y}$-component of the minimal distortion shift equation, the 
Hamiltonian constraint, and the maximal slicing condition.  Since we
require the values of the IRE to approach zero as 
${\cal{O}} ({\Delta \hat{x}}^2)$, we multiply the medium resolution
IRE ($\Delta \hat{x} = 0.03$) by a factor of 4, and we multiply the high
resolution IRE ($\Delta \hat{x} = 0.015$) by a factor of 16.  In this
way, it is simple to see if our numerical solver is consistent with
the IRE (and thus consistent with the original differential equations):
simply plot the values of the IREs scaled with the appropriate number
(1,4, and 16 for the low, medium, and high resolutions, respectively)
and see if the values are the same.  If they are the same, we are assured that
finite difference equations we are solving are consistent with the 
differential equations we want to solve, thus validating the code
as described above.

In observing the values of the IREs, we note that the residual appears
not to be converging to zero as the second power of the discretization
parameter $\Delta \hat{x}$ near the black hole, $\hat{x}_{BH} = -0.244$.
This is to be expected, as the conformal factor has a $1/\hat{r}$ pole
at that point, and higher order terms in the truncated Taylor series 
for the finite difference approximations are of the same order (or larger than)
the second-order term near this pole.  It could also be the case that
higher order terms in the truncated Taylor series for the finite 
difference approximations of
the $\hat{y}$-component of both the momentum constraints and the
minimal distortion shift equations (Figure~\ref{fig:ire1}) are causing
the lack of strict second-order convergence observed at the surface
of the neutron star ($\hat{x} = 0.4668$ and $1.0$);  both equations have
source terms with a discontinuous derivative.  It also could be the
case that setting stronger tolerances for the multigrid elliptic solver
would eliminate this behavior at these resolutions; we would normally
terminate the multigrid elliptic solve W-cycle~\cite{BrandtA94} when
the $L_2$-norm of the residual fell below 5,000 times the 
$L_2$-norm of the truncation error.  

Other than these noted deviations, we see strict second-order
convergence of the IREs.  This makes us confident that we have 
eliminated all errors in the code that could possibly cause the
solutions to the difference equations to converge to anything other
than solutions to the differential equations.

\begin{figure}
\epsfxsize=3in
\begin{center}
\leavevmode
\epsffile{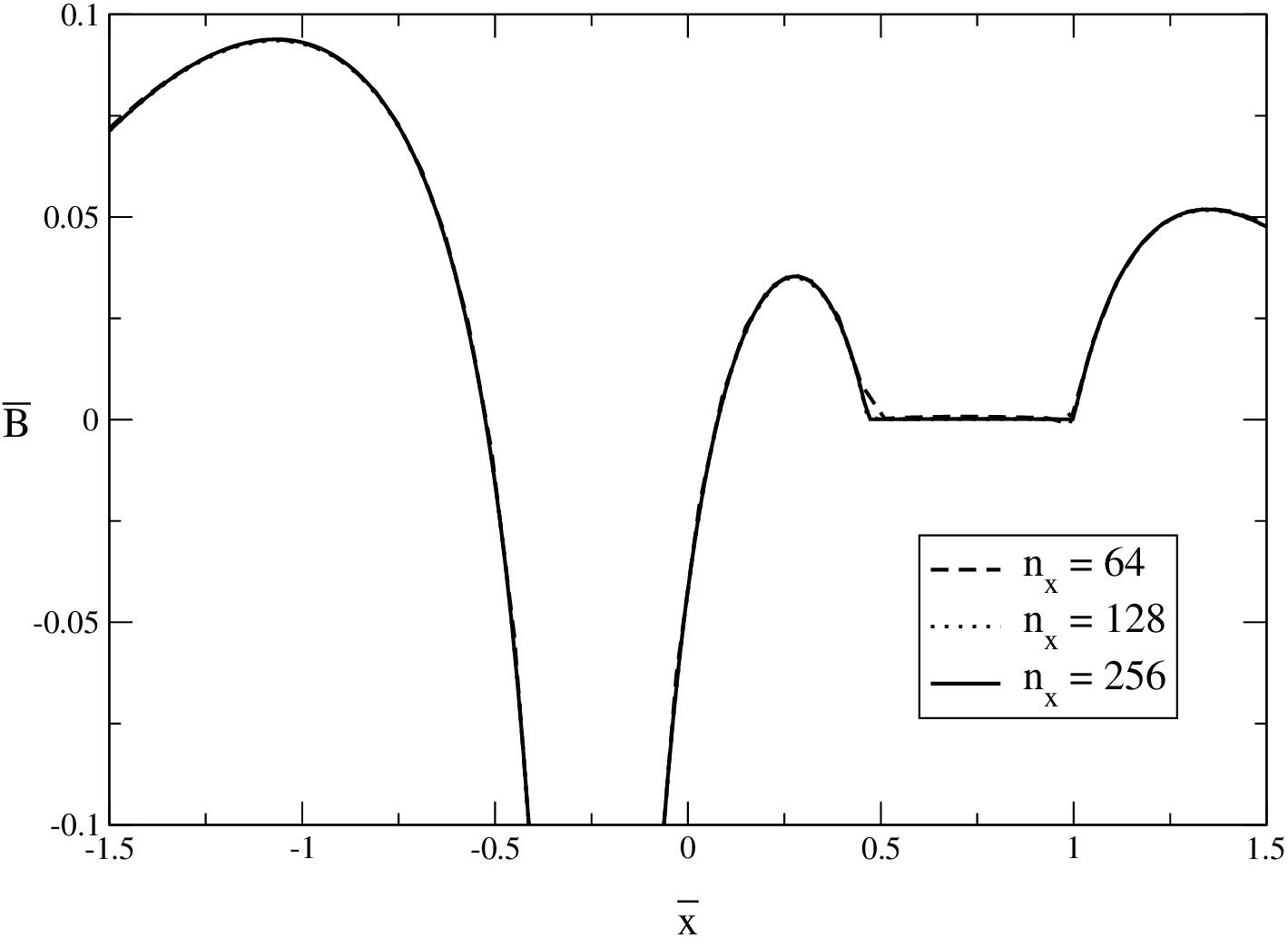}
\end{center}
\caption{ An independent residual evaluator (IRE) evaluated on the numerical
solution obtained for the configuration specified by ${\mu}_{BH} = 3$,
${\hat{x}}_A = 0.4668$, and ${({\rho}_0)}_{max} = 0.09265$.
Shown is the independent residual of Bernoulli's equation,
Eq.~\ref{eq:bernoulli}, along the $\hat{x}$-axis for
3 separate resolutions.  We only require this equation to be 
satisfied where ${\rho}_0 \neq 0$, namely, for $0.4668 < \hat{x} < 1.0$
(see Figure~\ref{fig:bernoulli2}).
}
\label{fig:bernoulli1}
\end{figure}

\begin{figure}
\epsfxsize=3in
\begin{center}
\leavevmode
\epsffile{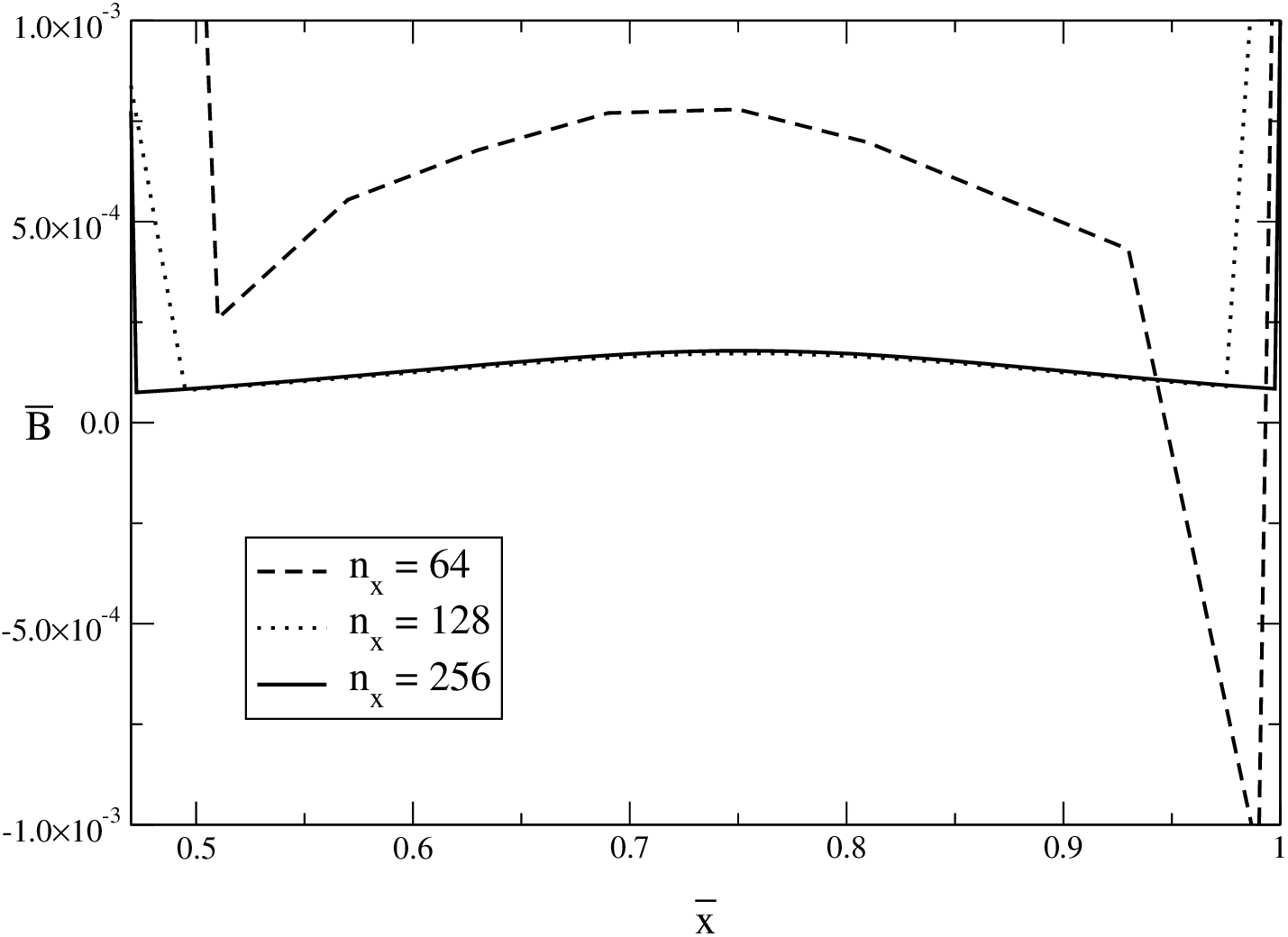}
\end{center}
\caption{ An independent residual evaluator (IRE) evaluated on the numerical
solution obtained for the configuration specified by ${\mu}_{BH} = 3$,
${\hat{x}}_A = 0.4668$, and ${({\rho}_0)}_{max} = 0.09265$.
Shown is the independent residual of Bernoulli's equation,
Eq.~\ref{eq:bernoulli}, along the $\hat{x}$-axis for
3 separate resolutions.  Only the $\hat{x}$ values where
${\rho}_0 \neq 0$ are shown ($0.4668 < \hat{x} < 1.0$).
}
\label{fig:bernoulli2}
\end{figure}

For completeness, we also show an IRE for Bernoulli's equation,
Eq~\ref{eq:bernoulli}, in 
Figures~\ref{fig:bernoulli1}~and~\ref{fig:bernoulli2}.  Although
it is not a differential equation, and thus we do not expect
to see convergence, we see that the algorithm described in
Section~\ref{sec:Algorithm} is solving the relativistic
Bernoulli's equation to almost 1 part in 10,000 inside the
neutron star, $0.4668 < \hat{x} < 1.0$.

\section{Sequences of constant NS baryonic mass, constant BH bare
mass BH/NS binaries}
\label{sec:Sequences}

Now that we have successfully validated a code that can produce 
fully general relativistic initial data that corresponds to a neutron
star and a black hole in quasiequilibrium, quasicircular orbit, we are ready
to use this as initial data for a numerical evolution code that can
handle both black holes and general relativistic hydrodynamics.  
However, if we choose to numerically evolve
an initial data configuration that is too far
away from the innermost stable circular orbit (ISCO), we will have to
numerically follow many orbital periods before the final plunge, which would, 
at best, be a difficult task.  Conversely, if we choose to numerically
evolve an initial data 
configuration that is too far inside the ISCO, we will be starting
with non-physical initial data.  As stated in the abstract of this 
paper, the only way to truly find the ISCO of a particular system will
be through fully general relativistic numerical studies.  However,
we can hope to find an approximate location of the ISCO by 
following~\cite{Baumgarte98b,Mathews00,Gourgoulhon01,Cook94,Baumgarte00a} 
and studying sequences of initial data sets that have both constant
baryonic mass and constant black hole mass.
We will define an effective
binding energy per unit rest mass, $E_b$, 
in terms of the ADM mass of the NS ($M_{NS}$), the bare
mass of the BH ($M_{BH}$), and the total ADM mass of the system $M_{ADM}$.
We will take the minimum (when it exists) of this effective 
binding energy $E_b$ to be the configuration which approximates
the ISCO configuration of the system.  Fixing the rest mass of the
system and the binding energy per unit mass in this way leaves one
free parameter left from the initial 3 free parameters ($\mu_{bn}$,
${\hat{x}}_A$, and ${({\rho}_0)}_{max}$).  We can therefore find
ISCO configurations for varying values of the ratio of the BH and
NS mass, $\mu_{bn}$.

Obviously,
the baryon number in the NS will be conserved during a quasiequilibrium
orbit of a NS and BH.  It may be argued, however, that one should fix
a different mass (e.g., the apparent horizon mass) for the black hole
when looking at sequences to try to find the ISCO.  Technically, there
is no exactly conserved BH mass during the quasiequilibrium orbit of a
NS and BH; even the event horizon mass will not be strictly conserved
during the quasiequilibrium orbits down to the ISCO.  
Here, we find that holding
the bare mass of the BH fixed to be equivalent (to within numerical
accuracy) 
to holding the apparent horizon mass fixed.  For small values of
$\mu_{bn}$, this is due to the fact that the NS is very far away
from the BH compared to the mass of the BH and thus does not much
affect the area of the apparent horizon.  For large values of
$\mu_{bn}$, while the NS is closer to the BH, the mass of the NS does
not much affect the area of the apparent horizon.  

We define the binding energy per unit rest mass as
\begin{equation}
E_b = \frac {M_{ADM} - M_{BH} - M_{NS}} { {(M_0)}_{NS}}
\end{equation}
where $M_{BH}$ is the bare mass of the BH, ${(M_0)}_{NS}$ is the 
rest mass of the NS, $M_{NS}$ is the ADM mass of a NS in isolation with
a rest mass of ${(M_0)}_{NS}$, and $M_{ADM}$ is the ADM mass of the entire
system, defined to be 
\begin{equation}
M_{ADM} = \frac {1}{16 \pi} \lim_{r \rightarrow \infty} 
\sum^{3}_{i, j = 1} \oint \; dA^{i} \; \left (
\frac {\partial {\gamma}_{ij}}{\partial x^j} - 
\frac {\partial {\gamma}_{jj}}{\partial x^i} \right )
\end{equation}
where the integration surface is of constant radial coordinate $r$.  Using
Stoke's theorem, along with our particular form of the conformal factor
(Eq.~\ref{eq:conf_fact})
of our conformally flat 3-metric,
we can write this integral for the ADM mass of the system as 
\begin{equation}
M_{ADM} = M_{BH} - \frac {1} {2 \pi} \int d^3x \; {\partial}^i
{\partial}_i \chi
\end{equation}
The integrand ${\partial}^i {\partial}_i \chi$ can be replaced using the 
Hamiltonian constraint, Eq.~\ref{eq:ham_nohat}, in terms of the
conformal extrinsic curvature, conformal factor, and matter
source terms.  It is in this form that we numerically integrate (in 
hatted coordinates ${\hat{x}}^i$) to obtain the ADM mass of the system. 

Recall that, for any given mass ratio parameter ${\mu}_{bn}$, our
numerical algorithm described in Section~\ref{sec:Algorithm} does not
allow us to fix the rest mass of the NS.  Instead, the remaining freedom
of the system is parameterized in the numerical algorithm
by the separation parameter
${\hat{x}}_A$ and the maximum rest mass density ${({\rho}_0)}_{max}$
of the NS. In order to construct sequences of constant rest mass
initial data for a particular mass ratio ${\mu}_{bn}$, 
we must compute initial data sets for various
values of parameters ${\hat{x}}_A$ and ${({\rho}_0)}_{max}$.  By interpolating
these results, we can find constant rest mass sequences.  Specifically,
we chose various values (five, typically) for the separation parameter
${\hat{x}}_A$.  Then, for each choice of ${\hat{x}}_A$ we solve for the
initial data set corresponding to a series of different parameter values
of ${({\rho}_0)}_{max}$ and connect these results using a cubic spline
(whose independent parameter is the rest mass of the system).  We can then,
for each value of ${\hat{x}}_A$, find the details of any initial
data set determined by any particular value of the rest mass.  In this
way, we can plot the binding energy per unit rest mass $E_b$ as
a function of angular frequency parameter $\Omega$.

\begin{figure}
\epsfxsize=3in
\begin{center}
\leavevmode
\epsffile{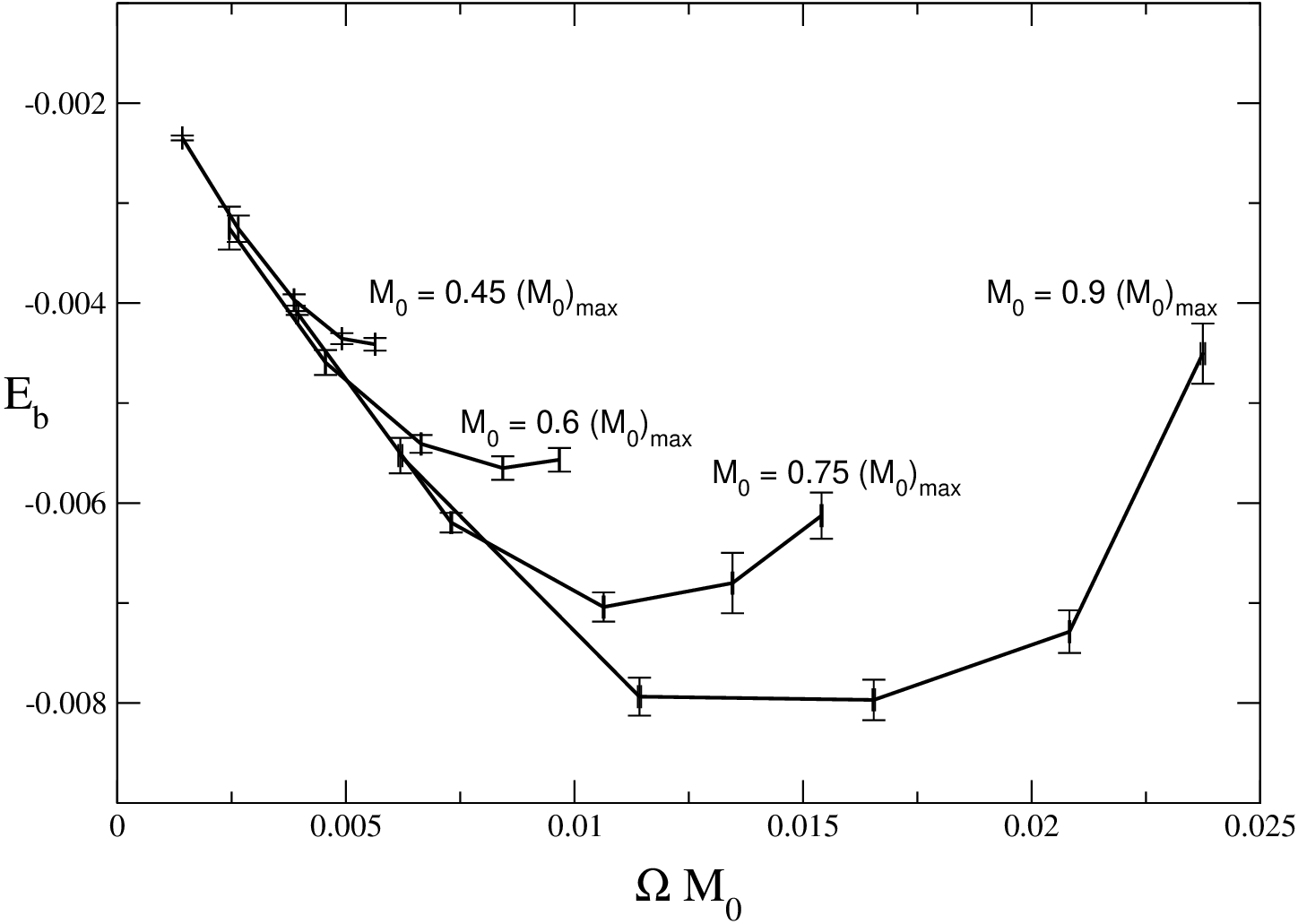}
\end{center}
\caption{ The binding energy per unit rest mass vs. the 
angular velocity (scaled by the rest mass of the system) for
constant rest mass sequences where the ratio of the BH mass
to the NS mass, ${\mu}_{bn}$, is 1.  Shown are sequences whose 
rest mass are $0.45$, $0.6$, $0.75$, and $0.9$ times the
maximum stable rest mass of a NS for a $\Gamma = 2$
polytropic equation of state.
In units where $G=c=k=1$, the maximum rest mass for a $\Gamma = 2$
polytropic star is ${(M_0)}_{max} = 0.179862$.  The minimum of each
sequence (when it exists) is taken to be an approximate ISCO 
configuration (see Tables~\ref{tab:isco1}~and~\ref{tab:isco2}).
}
\label{fig:mu1_eb}
\end{figure}

In Figure~\ref{fig:mu1_eb} we plot, for mass ratio
parameter ${\mu}_{bn} = 1$,  the binding energy per unit mass $E_b$
for a series of constant rest mass sequences.  We consider the 
minimum of each sequence to be approximations to the ISCO configuration
for the binary BH/NS system with the given rest mass.
Note that no minimum exists for the 
$M_0 = 0.45 {(M_0)}_{max}$ sequence in Figure~\ref{fig:mu1_eb}.  We
do not claim that a ISCO configuration does not exist for
a BH/NS system where $M_0 = 0.45 {(M_0)}_{max}$. Instead,
this is a result of the fact that our numerical algorithm
described in  Section~\ref{sec:Algorithm} does not converge
on a solution when ${\hat{x}}_A < 0$ (i.e., when the center of mass
of the system is inside the NS).  Note
from Table~\ref{tab:configs} that the smallest value of
${\hat{x}}_A$ used in the construction of Figure~\ref{fig:mu1_eb}
is, in fact, ${\hat{x}}_A = 0$.  These points correspond to the
highest value of $\Omega$ for each constant rest mass sequence in
Figure~\ref{fig:mu1_eb}.  These points are such that the
center of mass of the system exists at the surface of the NS.

\begin{table}
\begin{tabular}{|c|c|c||c|}  \hline \hline
\hspace{0.0cm} {\Large ${\mu}_{bn}$} \hspace{0.0cm}  &
\hspace{0.0cm} {\Large $n_{\hat{x}}$} \hspace{0.0cm} &
\hspace{0.0cm} {\Large $\Delta\hat{x}$} \hspace{0.0cm} &
{\Large ${\hat{x}}_A$}  \\ \hline \hline
1  & 256 & 0.015 & 0.0, 0.08397, \\   \cline{1-3}
1  & 128 & 0.03  & 0.1881, 0.3215, \\ \cline{1-3}
1  & 160 & 0.015 & 0.5 \\ \hline
3  & 256 & 0.015 & 0.3, 0.3771,  \\   \cline{1-3}
3  & 128 & 0.03  & 0.4668, 0.5727, \\ \cline{1-3}
3  & 160 & 0.015 & 0.7 \\ \hline
10 & 256 & 0.01     & 0.5, 0.5796,    \\   \cline{1-3}
10 & 128 & 0.02     & 0.6707, 0.7761, \\   \cline{1-3}
10 & 160 & 0.014375  & 0.8358, 0.9 \\ \hline \hline
\end{tabular}
\vspace{3.0mm}
\caption{ Configurations used to estimate the truncation errors
and boundary errors of the constant rest mass,
constant BH mass sequences in 
Figures~\ref{fig:mu1_eb},~~\ref{fig:mu3_eb},~and~\ref{fig:mu10_eb}
and of the ISCO configurations 
of Tables~\ref{tab:isco1}~and~\ref{tab:isco2} (see discussion).
}
\label{tab:configs}
\end{table}

Observe the error bars in Figure~\ref{fig:mu1_eb}.
It seems that every branch of science except numerical relativity
has, in the past, at least ${\it tried}$ to quantify the errors
in any given measurement and/or calculation.  While in the past it
may have been true that any reasonable measure of error in a 3D numerical
relativity calculation would result
in error bars of the same magnitude of (or larger than) the
actual quantity being 
calculated, this is no longer the case.  Available computer size and power
are now such that we can make reasonable attempts to
quantify the errors in our numerical relativity calculations.  In our
case, there are two sources of numerical approximation errors.
The first source of numerical error is usually
referred to as 'truncation error', and is the error induced by 
neglecting higher order terms in the Taylor series expansion of our 
finite difference approximations of derivatives of functions.  In our
case, we have used second order differencing, which means the 
highest order non-zero truncation error is of order ${\Delta\hat{x}}^2$, i.e., 
the truncation error goes to zero as ${\Delta\hat{x}}^2$ goes to zero.
The second source of numerical error is due to the fact that our computational
outer boundary (which is cubical) is at a finite distance from the origin,
whereas the boundary conditions to our elliptic equations are formulated
in terms of the behavior of the fields at $\infty$
(e.g., $(\phi-1) \rightarrow {\cal{O}}(1/\hat{r})$ as $\hat{r} \rightarrow \infty$).
We use robin boundary conditions in our multigrid elliptic solver.  
E.g., we could place the condition $(\phi - 1) \sim (1/\hat{r})$ on the 
field $\phi$ at a finite distance from the origin of our coordinate
system.  In doing so, we neglect higher order terms in a $1/\hat{r}$ expansion
of $\phi$ at the boundary.  Thus, the error we make by placing our 
computational boundary at a finite distance ${\hat{r}}_c$ from the
origin is of order ${\cal{O}}(1/{\hat{r}_c}^2)$.  Therefore, the total
error made in the computation of any quantity $Q$ in our numerical
method can be modeled as the following relationship between the
exact value $Q_e$ (the exact value one would obtain if one solved the
differential equations directly) and the quantity obtained by numerical 
calculation $Q_n$ as
\begin{equation}
Q_n = Q_e + e_t {\Delta\hat{x}}^2 + e_b \frac {1}{ {\hat{r}_c}^2} + \cdots,
\end{equation}
where $e_t$ and $e_b$ are constants and ``$\cdots$'' represents
the higher order contributions to the error.
By performing a numerical calculation with 3 different sets of
resolution/gridpoints, we can solve for the unknown quantities
$Q_e$, $e_t$, and $e_b$.  We now have a measure of the numerical error for 
our best calculation $Q_n$, namely, the calculation with the highest resolution
and/or with computational boundary furthest from the origin.  For the
numerical calculations performed for Figure~\ref{fig:mu1_eb}
(see Table~\ref{tab:configs} for the numerical 
configurations used in the calculation
of the errors), as well as for all of the numerical results
presented in this paper, we attach errors that are equal to 
\begin{equation}
{(\Delta Q)}_{error} = max\{|Q_n - Q_e|, |e_t {\Delta\hat{x}}^2|,
|e_b \frac {1}{ {\hat{r}_c}^2}|\}
\label{eq:errorbar}
\end{equation}
where $Q_n$, $\Delta\hat{x}$, and $\hat{r}_c$ is the result and 
configuration of the highest resolution run (the configuration with
$n_{\hat{x}} = 256$ in Table~\ref{tab:configs});  $Q_e$, 
$e_t$, and $e_b$ are computed (separately for each quantity $Q$)
as described above.  Note that, while this prescription 
(Eq.~\ref{eq:errorbar}) for the
size of the error bars can overestimate the error, it will never 
underestimate the magnitude of the largest error terms
in either the truncation error or the boundary error.

\begin{figure}
\epsfxsize=3in
\begin{center}
\leavevmode
\epsffile{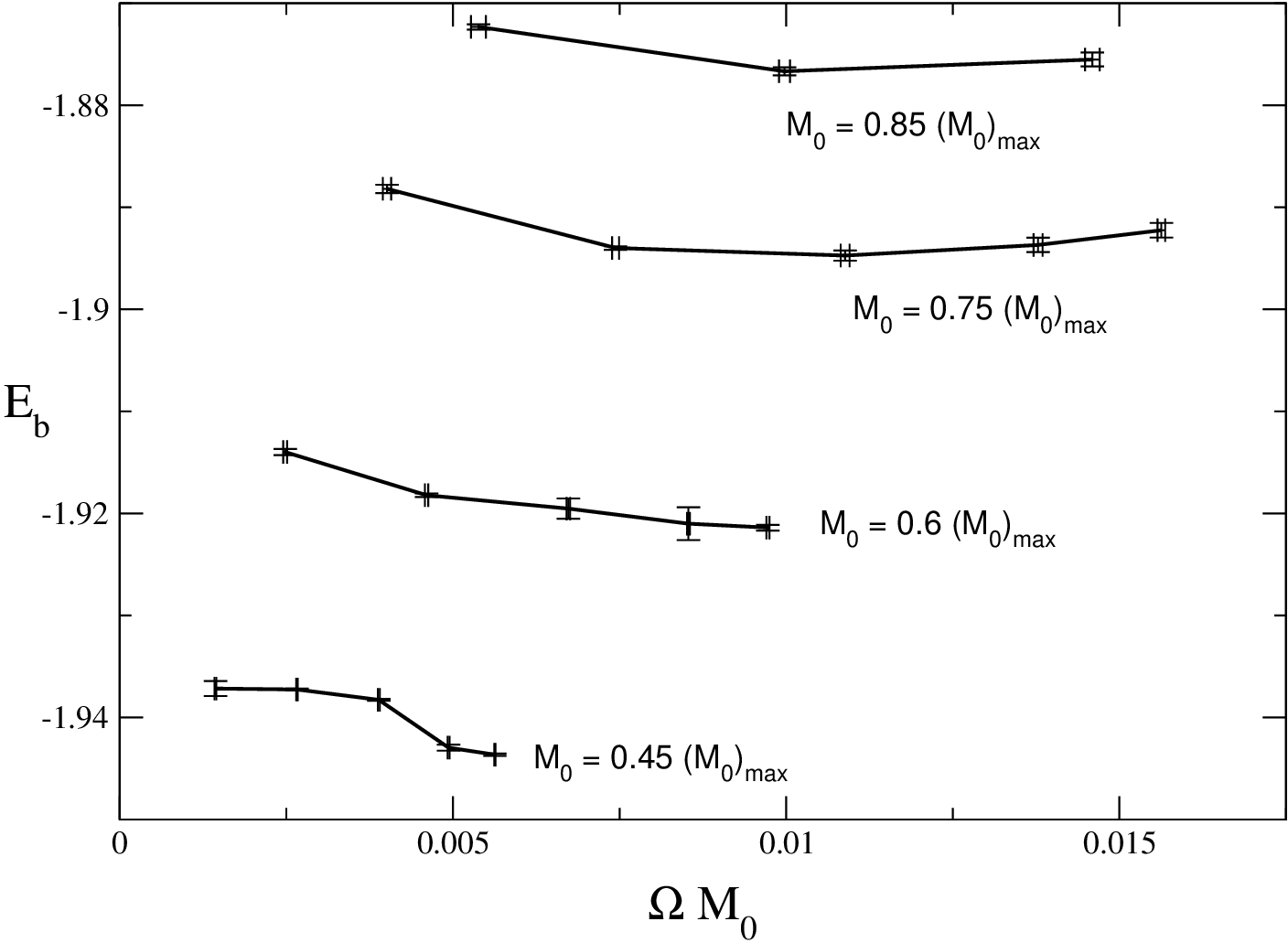}
\end{center}
\caption{ The binding energy per unit rest mass vs. the
angular velocity (scaled by the rest mass of the system) for
constant rest mass sequences where the ratio of the BH mass
to the NS mass, ${\mu}_{bn}$, is 3.  Shown are sequences whose
rest mass are $0.45$, $0.6$, $0.75$, and $0.85$ times the
maximum stable rest mass of a NS for a $\Gamma = 2$
polytropic equation of state.
In units where $G=c=k=1$, the maximum rest mass for a $\Gamma = 2$
polytropic star is ${(M_0)}_{max} = 0.179862$.  The minimum of each
sequence (when it exists) is taken to be an approximate ISCO
configuration (see Tables~\ref{tab:isco1}~and~\ref{tab:isco2}).
}
\label{fig:mu3_eb}
\end{figure}

\begin{figure}
\epsfxsize=3in
\begin{center}
\leavevmode
\epsffile{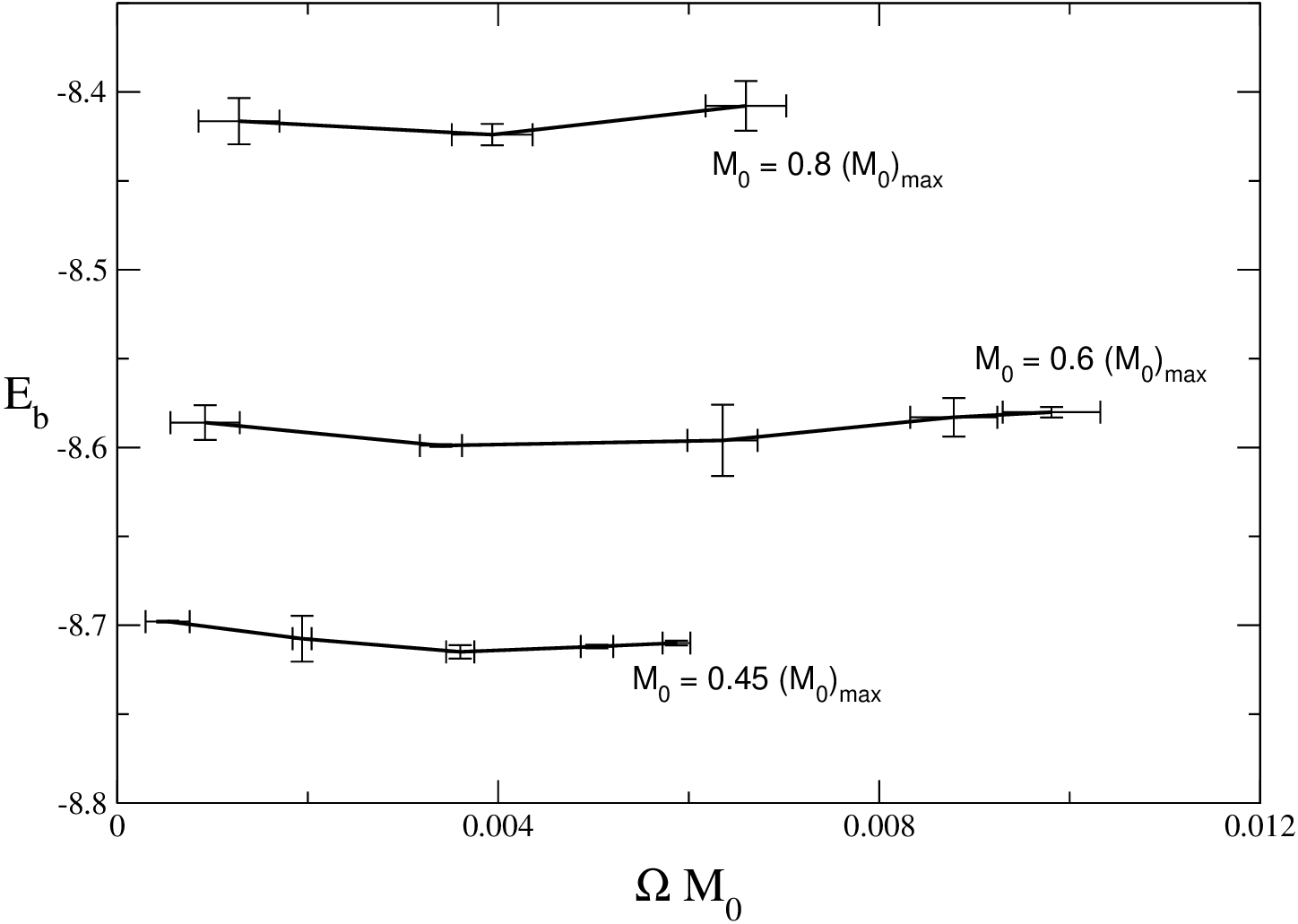}
\end{center}
\caption{The binding energy per unit rest mass vs. the
angular velocity (scaled by the rest mass of the system) for
constant rest mass sequences where the ratio of the BH mass
to the NS mass, ${\mu}_{bn}$, is 10.  Shown are sequences whose
rest mass are $0.45$, $0.6$, and $0.8$ times the
maximum stable rest mass of a NS for a $\Gamma = 2$
polytropic equation of state.
In units where $G=c=k=1$, the maximum rest mass for a $\Gamma = 2$
polytropic star is ${(M_0)}_{max} = 0.179862$.  The minimum of each
sequence (when it exists) is taken to be an approximate ISCO
configuration (see Tables~\ref{tab:isco1}~and~\ref{tab:isco2}).
}
\label{fig:mu10_eb}
\end{figure}

In Figures~\ref{fig:mu3_eb}~and~\ref{fig:mu10_eb}, we show the
binding energy per unit mass $E_b$ for a series of constant
rest mass sequences for mass ratio ${\mu}_{bn} = 3$ and $10$, 
respectively.  Note from Table~\ref{tab:configs} that the minimum 
separation parameter ${\hat{x}}_A$ used for the 
${\mu_{bn}} = 3$ and ${\mu_{bn}} = 10$ calculations are
${\hat{x}}_A = 0.3$ and ${\hat{x}}_A = 0.5$, respectively.  We have
found that, for lower values of ${\hat{x}}_A$ (i.e., smaller
separation), the Roche limit~\cite{Kopal78} is reached
for large values of ${(\rho_0)}_{max}$.  This result is
probably highly dependent on the EOS (we use an adiabatic
index $\Gamma = 2$ here).  However, as can be seen from 
Figures~\ref{fig:mu1_eb},~\ref{fig:mu3_eb},~and~\ref{fig:mu10_eb}, a minimum
in binding energy per unit rest mass $E_b$ is found before the 
Roche limit is reached, at least for higher rest mass cases.
Notice also that the maximum rest mass reported for 
Figures~\ref{fig:mu1_eb},~\ref{fig:mu3_eb},~and~\ref{fig:mu10_eb}
(corresponding to ${\mu_{bn}} = 1,3$, and $10$, respectively) is
$0.9 {(M_0)}_{max}$, $0.85 {(M_0)}_{max}$, and $0.8 {(M_0)}_{max}$,
respectively.  We have found that, in each case, for
slightly higher rest mass configurations, the numerical algorithm
described in Section~\ref{sec:Algorithm} fails to converge
to a simultaneous solution in a reasonable amount of iterations.  
However, since we are really only interested in initial data for future
numerical simulations, a maximum rest mass of the NS in the range of
$0.8 {(M_0)}_{max}$ - $0.9 {(M_0)}_{max}$ is certainly adequate.

Note that the errors for the binding 
energy per unit mass $E_b$ for the ${\mu}_{bn} = 10$ case 
(Figure~\ref{fig:mu10_eb}) are over an order of magnitude
larger than the errors in the ${\mu}_{bn} = 1$ case
(Figure~\ref{fig:mu1_eb}).  This is due to the fact that the relative
separation is much larger in the higher ${\mu}_{bn}$ case
(the ${\hat{x}}_A$ used for ${\mu}_{bn} = 10$ is closer
to 1, see Table~\ref{tab:configs});  more 
resolution would be needed to get the same number of points across 
the NS.  

In Tables~\ref{tab:isco1}~and~\ref{tab:isco2}, 
we tabulate the results of the approximate
ISCO configurations which we define to be the minimum of the 
binding energy per unit mass $E_b$ for a constant rest mass 
sequence (Figures~\ref{fig:mu1_eb}~-~\ref{fig:mu10_eb}).  To find the 
minimum, we fit a second order polynomial to the minimum 
data point, e.g. in Figure~\ref{fig:mu1_eb}, and its nearest
neighbor on either side.  The minimum of this polynomial designates
a good approximation of the minimum of the actual curve.  The
errors in Tables~\ref{tab:isco1}~and~\ref{tab:isco2} are computed 
using the same method as that used to calculate the error
bars in Figures~\ref{fig:mu1_eb}~-~\ref{fig:mu10_eb} (see
Eq.~\ref{eq:errorbar}).   Note that the relative
errors of the tabulated values increase for larger values of the 
BH/NS mass ratio parameter ${\mu}_{bn}$.  This is due to the fact that
as the mass of the BH increase relative to the mass
of the NS, the separation between the two bodies for the ISCO
configuration
increases in units of the NS radius.
As a result, there is
less numerical resolution per NS radius for higher ${\mu}_{bn}$,
resulting in a relative loss of accuracy. 
For configurations where ${\mu}_{bn} > 10$, it is likely that
mesh refinement techniques will be need to accurately 
resolve both the NS and BH.

\onecolumn

\begin{table}
\begin{tabular}{|c|c|c|c|c|c|}  \hline \hline
\hspace{0.0cm} {\Large ${\mu}_{bn}$} \hspace{0.0cm}  &
\hspace{0.0cm} {\Large $M_0/{(M_0)}_{max}$} \hspace{0.0cm}  &
\hspace{0.0cm} {\Large $E_b$} \hspace{0.0cm}  &
\hspace{0.0cm} {\Large $J/({M_0}^2)$} \hspace{0.0cm} &
\hspace{0.0cm} {\Large $\Omega M_0$} \hspace{0.0cm}  &
\hspace{0.0cm} {\Large ${(\rho_0)}_{max}$} \hspace{0.0cm}  \\ \hline \hline
{ $1$} & 
   { $0.6$} & 
   { $-0.00283\pm0.00012$} & 
   { $2.9069\pm0.0033$}  &
   { $0.00855\pm0.00020$} &
   { $0.06326\pm0.00018$} \\ \hline
{ $1$} &      
   { $0.75$} &      
   { $-0.00353\pm0.00025$} &      
   { $2.609\pm0.025$}  &
   { $0.01127\pm0.00099$} &
   { $0.09589\pm0.00061$} \\ \hline
{ $1$} &      
   { $0.9$} &     
   { $-0.00403\pm0.00024$} &        
   { $2.3887\pm0.0039$}  &
   { $0.01417\pm0.00038$} &
   { $0.14947\pm0.00083$} \\ \hline
{ $3$} &      
   { $0.75$} &      
   { $-0.94738\pm0.00053$} &        
   { $6.396\pm0.059$}  &
   { $0.0103\pm0.0013$} &
   { $0.09691\pm0.00098$} \\ \hline
{ $3$} &      
   { $0.85$} &     
   { $-0.93846\pm0.00060$} &        
   { $6.138\pm0.017$}  &
   { $0.0113\pm0.0010$} &
   { $0.1284\pm0.0015$} \\ \hline
{ $10$} &      
   { $0.45$} &     
   { $-4.3575\pm0.0028$} &        
   { $21.5\pm1.6$}  &
   { $0.0038\pm0.0019$} &
   { $0.0424\pm0.0020$} \\ \hline
{ $10$} &     
   { $0.6$} &     
   { $-4.300\pm0.015$} &         
   { $19.9\pm1.4$}  &
   { $0.0044\pm0.0041$} &
   { $0.0657\pm0.0017$} \\ \hline
{ $10$} &     
   { $0.8$} &      
   { $-4.212\pm0.034$} &        
   { $20.6\pm3.1$}  &
   { $0.0038\pm0.0071$} &
   { $0.1183\pm0.0061$} \\ \hline \hline
\end{tabular}
\vspace{0.0mm}
\caption{ ISCO configuration state parameters.  Results are
obtained by minimizing the binding energy per unit rest mass $E_b$ 
for constant rest mass sequences
(see Figures~\ref{fig:mu1_eb},~\ref{fig:mu3_eb},~and~\ref{fig:mu10_eb})
at the highest numerical resolution ($n_x = 256$).  Numerical errors
are calculated by minimizing $E_b$ for other numerical
configurations (see Table~\ref{tab:configs}) and calculating the
error as per Eq.~\ref{eq:errorbar}.  Shown are tabulated values 
for the ISCO configurations
of the binding energy per unit mass $E_b$, the total angular
momentum of the system $J$, the orbital angular velocity
parameter $\Omega$, and the maximum rest mass density 
of the NS ${(\rho_0)}_{max}$ for various choices of BH/NS mass 
ratio parameter ${\mu}_{bn}$ and total NS rest mass $M_0$.
}
\label{tab:isco1}
\end{table}

\begin{table}
\begin{tabular}{|c|c|c|c|c|c|}  \hline \hline
\hspace{0.0cm} {\Large ${\mu}_{bn}$} \hspace{0.0cm}  &
\hspace{0.0cm} {\Large $M_0/{(M_0)}_{max}$} \hspace{0.0cm}  &
\hspace{0.0cm} {\Large ${\hat{x}}_A$} \hspace{0.0cm}  &
\hspace{0.0cm} {\Large $C_B$} \hspace{0.0cm} &
\hspace{0.0cm} {\Large $\sigma$} \hspace{0.0cm}  &
\hspace{0.0cm} {\Large $P^y_{BH}$} \hspace{0.0cm} 
    \\ \hline \hline
{ $1$} & 
   { $0.6$} & 
   { $0.076\pm0.012$} & 
   { $0.87244\pm0.00049$}  &
   { $2.215\pm0.014$} &
   { $-0.01288\pm0.00016$} \\ \hline
{ $1$} &      
   { $0.75$} &      
   { $0.164\pm0.033$} &      
   { $0.8342\pm0.0021$}  &
   { $2.113\pm0.059$} &
   { $-0.01797\pm0.00073$} \\ \hline
{ $1$} &      
   { $0.9$} &     
   { $0.2488\pm0.0098$} &        
   { $0.78917\pm0.00044$}  &
   { $2.000\pm0.012$} &
   { $-0.02344\pm0.00032$} \\ \hline
{ $3$} &      
   { $0.75$} &      
   { $0.484\pm0.037$} &        
   { $0.7886\pm0.0074$}  &
   { $3.18\pm0.19$} &
   { $-0.0362\pm0.0028$} \\ \hline
{ $3$} &      
   { $0.85$} &     
   { $0.540\pm0.021$} &        
   { $0.7605\pm0.0052$}  &
   { $3.16\pm0.11$} &
   { $-0.0424\pm0.0023$} \\ \hline
{ $10$} &      
   { $0.45$} &     
   { $0.66\pm0.14$} &        
   { $0.824\pm0.038$}  &
   { $5.6\pm1.2$} &
   { $-0.0274\pm0.0075$} \\ \hline
{ $10$} &     
   { $0.6$} &     
   { $0.73\pm0.12$} &         
   { $0.787\pm0.068$}  &
   { $5.459\pm0.058$} &
   { $-0.039\pm0.022$} \\ \hline
{ $10$} &     
   { $0.8$} &      
   { $0.85\pm0.18$} &        
   { $0.76\pm0.13$}  &
   { $9.8\pm3.2$} &
   { $-0.044\pm0.050$} \\ \hline \hline
\end{tabular}
\vspace{0.0mm}
\caption{ ISCO configuration state parameters.  Results are
obtained by minimizing the binding energy per unit rest mass $E_b$ 
for constant rest mass sequences
(see Figures~\ref{fig:mu1_eb},~\ref{fig:mu3_eb},~and~\ref{fig:mu10_eb})
at the highest numerical resolution ($n_x = 256$).  Numerical errors
are calculated by minimizing $E_b$ for other numerical
configurations (see Table~\ref{tab:configs}) and calculating the
error as per Eq.~\ref{eq:errorbar}.  Shown are tabulated values
for the ISCO configurations
of the separation parameter ${\hat{x}}_A$, the constant $C_B$
on the right hand side of Bernoulli's Eq.~\ref{eq:bernoulli_hat},
the scaling parameter $\sigma$, and the momentum of the BH
$P^y_{BH}$ (related to the value in the computational 
coordinates $\{ {\hat{x}}^i \}$ by 
$P^y_{BH} = \sigma {\hat{P}}^{\hat{y}}_{BH}$)
for various choices of BH/NS mass
ratio parameter ${\mu}_{bn}$ and total NS rest mass $M_0$.
}
\label{tab:isco2}
\end{table}

\twocolumn

The total angular momentum of the system about the center of mass
(which corresponds to the origin of our coordinate system) is
defined as~\cite{Wald84}
\begin{equation}
J^i_{ADM} = \frac {\epsilon^{ijk}}{8 \pi} \lim_{r \rightarrow \infty}
\oint \; dA^{m} \; \left ( x_j ( K_{km} - K \gamma_{km}) \right ),
\end{equation}
With our configuration, the $z$-component reduces to
\begin{equation}
J \equiv J^z_{ADM} = x_{BH} P^y_{BH} + \int \; d^3x \; {\phi}^{10}
{\rho_0}hW^2(xv^y - yv^x)
\label{eq:angmom}
\end{equation}
Note that the integral in Eq.~\ref{eq:angmom} has support only where
${\rho_0} \neq 0$, namely, inside the NS. 

\begin{figure}
\epsfxsize=3in
\begin{center}
\leavevmode
\epsffile{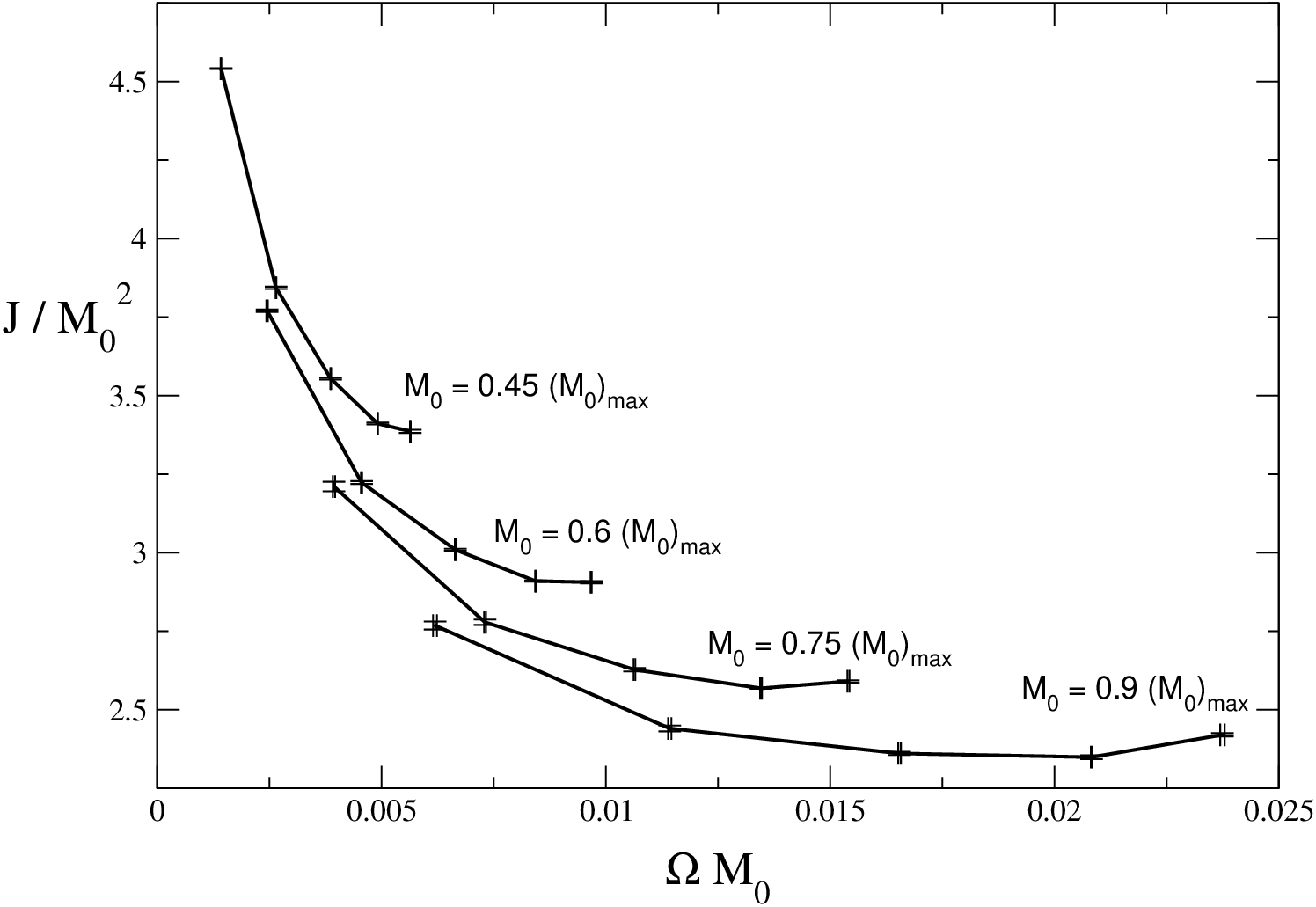}
\end{center}
\caption{The total angular momentum of the system
vs. the
angular velocity for
constant rest mass sequences where the ratio of the BH mass
to the NS mass, ${\mu}_{bn}$, is 1.  Shown are sequences whose
rest mass are $0.45$, $0.6$, $0.75$, and $0.9$ times the
maximum stable rest mass of a NS for a $\Gamma = 2$
polytropic equation of state.
In units where $G=c=k=1$, the maximum rest mass for a $\Gamma = 2$
polytropic star is ${(M_0)}_{max} = 0.179862$.
Plotted are the results from the highest resolution configuration
($n_x = 256$) from Table~\ref{tab:configs}, with error
bars signifying estimates of the numerical error as per
Eq.~\ref{eq:errorbar}.
}
\label{fig:mu1_j}
\end{figure}

\begin{figure}
\epsfxsize=3in
\begin{center}
\leavevmode
\epsffile{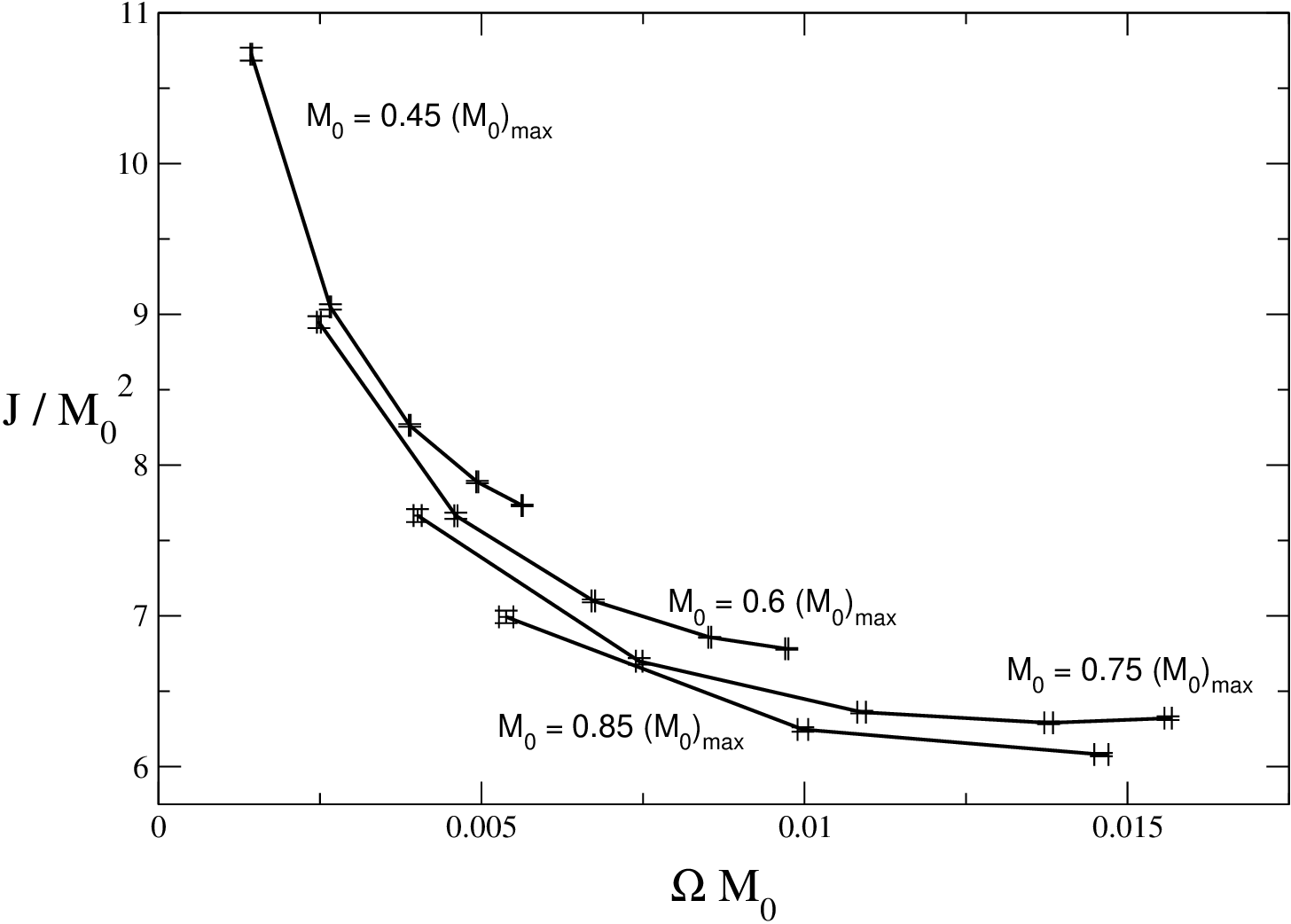}
\end{center}
\caption{The total angular momentum of the system
vs. the
angular velocity for
constant rest mass sequences where the ratio of the BH mass
to the NS mass, ${\mu}_{bn}$, is 3.  Shown are sequences whose
rest mass are $0.45$, $0.6$, $0.75$, and $0.85$ times the
maximum stable rest mass of a NS for a $\Gamma = 2$
polytropic equation of state.
In units where $G=c=k=1$, the maximum rest mass for a $\Gamma = 2$
polytropic star is ${(M_0)}_{max} = 0.179862$.  
Plotted are the results from the highest resolution configuration
($n_x = 256$) from Table~\ref{tab:configs}, with error
bars signifying estimates of the numerical error as per
Eq.~\ref{eq:errorbar}.
}
\label{fig:mu3_j}
\end{figure}

\begin{figure}
\epsfxsize=3in
\begin{center}
\leavevmode
\epsffile{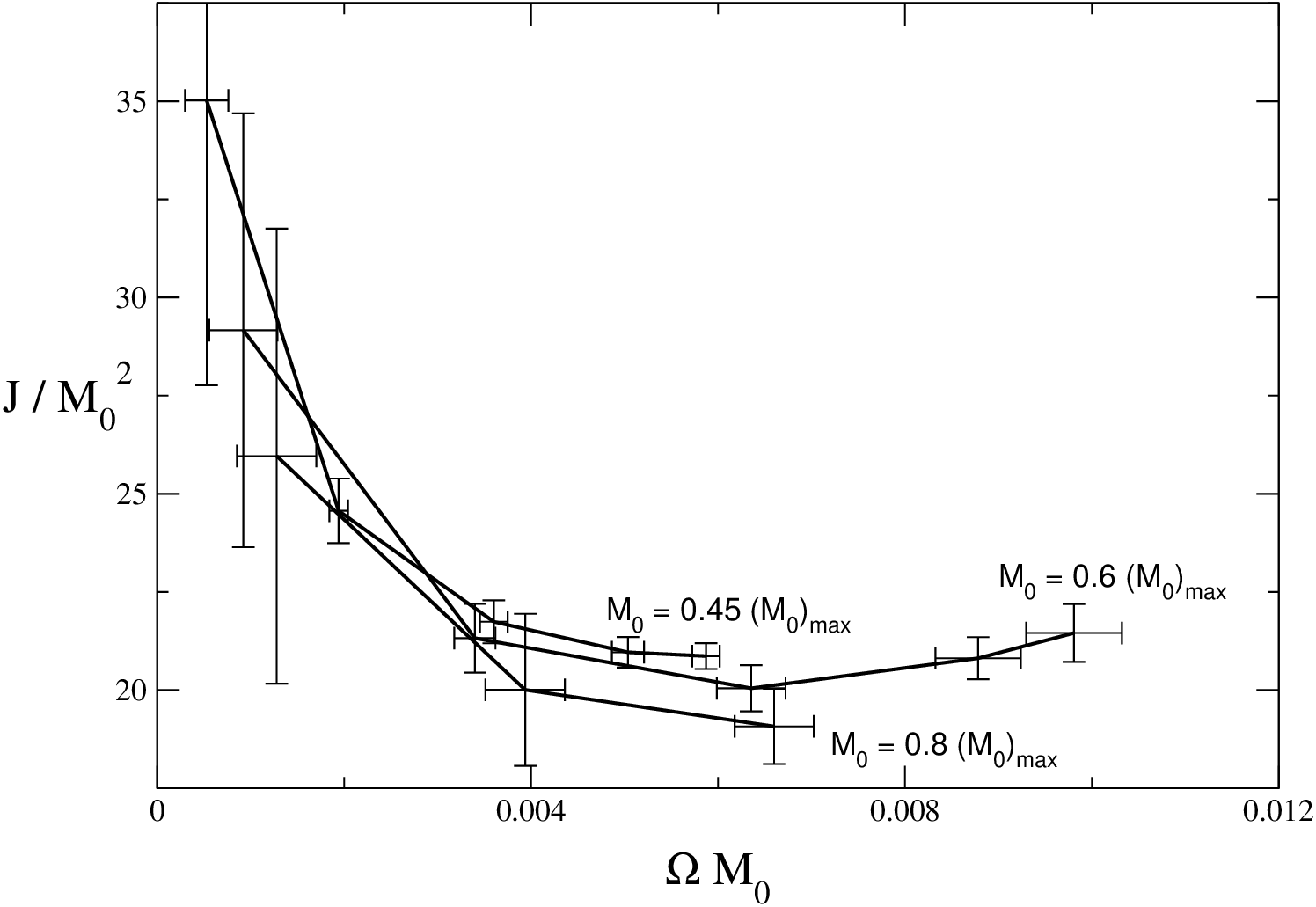}
\end{center}
\caption{The total angular momentum of the system
vs. the
angular velocity for
constant rest mass sequences where the ratio of the BH mass
to the NS mass, ${\mu}_{bn}$, is 10.  Shown are sequences whose
rest mass are $0.45$, $0.6$, and $0.8$ times the
maximum stable rest mass of a NS for a $\Gamma = 2$
polytropic equation of state.
In units where $G=c=k=1$, the maximum rest mass for a $\Gamma = 2$
polytropic star is ${(M_0)}_{max} = 0.179862$.  
Plotted are the results from the highest resolution configuration
($n_x = 256$) from Table~\ref{tab:configs}, with error
bars signifying estimates of the numerical error as per
Eq.~\ref{eq:errorbar}.
}
\label{fig:mu10_j}
\end{figure}

It is interesting to plot the total angular momentum of the system about the
center of mass, as was done for the binding energy per unit
mass in Figures~\ref{fig:mu1_eb}~-~\ref{fig:mu10_eb}.  In
Figures~\ref{fig:mu1_j}~-~\ref{fig:mu10_j}, we plot the unitless
quantity $J/M_0^2$ for values of mass ratio parameter
${\mu_{bn}} = 1,3,$ and $10$, respectively.  We use the same
numerical configurations as listed in Table~\ref{tab:configs}, and
compute the error bars as per Eq.~\ref{eq:errorbar}.  In~\cite{Baumgarte98b}
it is argued that, for constant rest
mass sequences,  the minimum of the angular momentum $J$ and the minimum of
the binding energy per unit mass $E_b$ should coincide, citing a
result~\cite{Bardeen73}
relating parameters of stationary solutions to the Einstein equations
not containing black holes, namely, $(dM_{ADM}) = \Omega \; (dJ_{ADM})$.  
In the study of quasiequilibrium 
corotating NS/NS binaries~\cite{Baumgarte98b}
it was shown that the minima of the binding energy and angular
momentum coincided to ``numerical accuracy''.
It could be asked if the configurations computed here share this 
feature.  For example, does the minimum of the binding
energy per unit mass $E_b$ for ${\mu}_{bn}=3$, $M_0/{(M_0)}_{max} = 0.75$
(see Figure~\ref{fig:mu3_eb}) correspond to the 
minimum of the of the angular momentum $J$ for the same case
(see Figure~\ref{fig:mu3_j})?  From Table~\ref{tab:isco1} we see
that this minimum corresponds to an orbital
angular frequency of $\Omega M_0 = 0.0103\pm0.0013$.  Performing a similar
calculation for the $M_0/{(M_0)}_{max} = 0.75$ case of Figure~\ref{fig:mu3_j},
computing numerical errors in the same fashion as for 
Tables~\ref{tab:isco1}~and~\ref{tab:isco2}, we obtain 
a value of orbital angular frequency $\Omega M_0 = 0.01376\pm0.00017$
at the minimum.  We see that the orbital angular frequency for the
minimum points do {\it not} agree within numerical errors.  The minimum
for the angular momentum occurs at a higher orbital angular
frequency (thus, at a smaller separation) than the minimum 
of the binding energy per unit mass.  In fact, for all of 
our configurations we find it to be true that the minimum of the 
angular momentum for a constant rest mass sequence occurs at
a smaller separation than the minimum of the binding
energy per unit mass.  

Why is this result different from 
previous studies~\cite{Baumgarte98b}?  There are many possible 
reasons, all of which could be contributing to some degree.  One
difference is that our configurations contain a black hole.
In this case, the result relating stationary solutions
to Einstein's equations~\cite{Bardeen73} is
\begin{equation}
dM_{ADM} = \frac {1}{8 \pi} \kappa_H \; dA_H + 
   \Omega_H \; dJ_H + \Omega \; dJ_{ADM} 
\label{eq:bh_cons}
\end{equation}
where $\kappa_H$ is the gravitational acceleration on the event horizon,
$A_H$ is the area of the event horizon, $\Omega_H$ is the angular
velocity of the BH, and $J_H$ is the angular momentum of the BH;  all of
these quantities are computed on the event horizon of the BH.  However,
it is certainly not clear whether Eq.~\ref{eq:bh_cons} even applies in
our case.  It was derived~\cite{Bardeen73} assuming a strictly axisymmetric
spacetime (e.g., an axisymmetric BH with matter distributed axisymmetrically),
whereas we have a nonspinning black hole located away from the center
of mass of the configuration.
Another difference is the ansatz used here for the extrinsic 
curvature; Eqs.~\ref{eq:a_define}~and~\ref{eq:mom_ansatz} is 
slightly different than that used in other NS/NS sequence 
studies~\cite{Baumgarte98b,Mathews00,Gourgoulhon01},  where the time
derivative of the conformal metric is set to zero.
In order to keep
the momentum constraints regular at the black hole puncture, we were
induced to use Eqs.~\ref{eq:a_define}~and~\ref{eq:mom_ansatz} as 
the form of the extrinsic curvature, noting that this ansatz reduces
to the usual form (the time derivative of the
conformal metric vanishes, Eq.~\ref{eq:dt0_mom_ansatz}) 
in the limit as $M_{BH} \rightarrow 0$.
Finally, we note that the present study most likely has a better 
numerical accuracy, as we have used a resolution that is 2 times
finer than in~\cite{Baumgarte98b}.  It is possible that with both
an increase in resolution and a better estimate of the numerical 
error, the study in~\cite{Baumgarte98b} would have detected a difference
between minima of the binding energy and minima of the total
angular momentum.

\section{Conclusion}
\label{sec:Conclusion}

In this paper, we have described a method to numerically calculate
general relativistic initial data configurations corresponding to a 
binary BH/NS in quasicircular orbit.  We construct a code to carry out
the calculation, and carefully validate the code using an independent
residual evaluator convergence test on each configuration solved for
this paper.  Assuming a $\Gamma = 2$ polytrope, we construct sequences of
constant rest mass, constant black hole mass initial data sets.  By minimizing
an effective binding energy, we find approximate ISCO configurations
for mass ratios of ${\mu_{bn}} \equiv M_{BH}/M_{NS} = 1, 3,$ and $10$.
A technique for monitoring the numerical error is presented and
used in reporting all numerical results.  

These ISCO configurations are only approximate ones in that the
constant rest mass, constant BH mass sequences constructed here
are not solutions to the full Einstein equations (although each
set solves the Hamiltonian and momentum constraints).  This is true
for all quasiequilibrium 
studies~\cite{Baumgarte98b,Mathews00,Gourgoulhon01,Cook94,Baumgarte00a}.  
Only a full numerical general relativistic treatment will be able to
quantify the errors in quasiequilibrium studies, which may be 
relatively large as a binary system approaches the ISCO.

In this study, we have assumed the NS to be corotating about the
center of mass of the binary system.  However, it has been
shown~\cite{Bildsten92} that a NS in orbit together with
a BH would need an unnaturally high viscosity to lock
the spin and orbital angular velocities. 
It would therefore be more realistic to perform the calculation
presented here with an 
irrotational~\cite{Bonazzola97,Teukolsky98,Shibata98} NS.  This
added feature, along with a realistic cold equation of state, will be
the subject of a future study.


\acknowledgements We would like to thank Matthew Miller and 
Wai-Mo Suen for encouragement and
helpful discussions.  This research
was supported by NSF (PHY 96-00507, PHY 99-79985, and MCA 93S025) 
and NASA (NCCS5-153).  Computations were performed at the Center for 
Scientific Parallel Computing at Washington University, St. Louis, the
National Center for Supercomputing Applications at the
University of Illinois at Urbana-Champaign, and at the 
Numerical Aerospace Simulation facility at NASA's Ames Research 
Center in Mountain View, California.


\bibliographystyle{prsty}


\end{document}